\documentclass[twocolumn,english,superscriptaddress,pre,nofootinbib]{revtex4-1}
\usepackage{lmodern}
\usepackage{lmodern}
\usepackage[T1]{fontenc}
\RequirePackage[latin1, utf8]{inputenc}

\usepackage{color}
\usepackage{xcolor}
\usepackage{babel}
\usepackage{amsmath}
\usepackage{amssymb}
\usepackage{graphicx}
\usepackage[unicode=true,pdfusetitle,
bookmarks=true,bookmarksnumbered=false,bookmarksopen=false,
breaklinks=false,pdfborder={0 0 1},backref=false,colorlinks=true]
{hyperref}
\hypersetup{
	pdfborderstyle=,urlcolor=orange,linkcolor=blue,citecolor=red}

\makeatletter

\usepackage{etoolbox}

\usepackage{bigints}

\usepackage{algorithm}
\usepackage{algpseudocode}

\usepackage{color}
\usepackage{babel}
\usepackage{units}

\usepackage[normalem]{ulem}
\usepackage{bigints}

\usepackage{comment}

\usepackage{tikz}
\usepackage{color}
\usepackage{dsfont}
\usepackage[cal=boondoxo]{mathalfa}
\usepackage{grffile}
\usepackage[caption=false]{subfig}

\catcode`,\active

\catcode`\,12

\newsavebox{\@brx}
\newcommand{\llangle}[1][]{\savebox{\@brx}{\(\m@th{#1\langle}\)}%
	\mathopen{\copy\@brx\kern-0.5\wd\@brx\usebox{\@brx}}}
\newcommand{\rrangle}[1][]{\savebox{\@brx}{\(\m@th{#1\rangle}\)}%
	\mathclose{\copy\@brx\kern-0.5\wd\@brx\usebox{\@brx}}}

\makeatletter
\def\maketitle{
	\@author@finish
	\title@column\titleblock@produce
	\suppressfloats[t]}
\makeatother

\newcommand{\w}{\boldsymbol{w}}
\newcommand{\p}{\boldsymbol{p}}
\newcommand{\x}{\boldsymbol{x}}
\newcommand{\s}{\boldsymbol{s}}

\DeclareMathOperator*{\argmax}{argmax}

\allowdisplaybreaks

\makeatother

\begin{document}
	\title{Sampling the space of solutions of an artificial neural network}
	\date{\today}
	
	\author{Alessandro Zambon}
	\email{alessandro.zambon@unimi.it}  
	\affiliation{Department of Physics, University of Milan and INFN, via Celoria 16, 20133 Milano, Italy}
	
	\author{Enrico M. Malatesta}
	\email{enrico.malatesta@unibocconi.it}
	\affiliation{Department of Computing Sciences and Bocconi Institute for Data Science and
		Analytics (BIDSA), Bocconi University, 20136 Milano, Italy}
	
	\author{Guido Tiana}
	\email{guido.tiana@unimi.it}
	\affiliation{Department of Physics, University of Milan and INFN, via Celoria 16, 20133 Milano, Italy}	
	
	\author{Riccardo Zecchina}
	\email{riccardo.zecchina@unibocconi.it}
	\affiliation{Department of Computing Sciences and Bocconi Institute for Data Science and
		Analytics (BIDSA), Bocconi University, 20136 Milano, Italy}
	
	\begin{abstract}
		
		The weight space of an artificial neural network can be systematically explored using tools from statistical mechanics. We employ a combination of a  hybrid Monte Carlo algorithm which performs long exploration steps, a ratchet-based algorithm to investigate connectivity paths,  and coupled replica models simulations to study subdominant flat regions.  Our analysis focuses on one hidden layer networks and spans a range of energy levels and constrained density regimes.
		Near the interpolation threshold, the low-energy manifold shows a spiky topology. 
		In the overparameterized regime, however, the low-energy manifold becomes entirely flat, forming an extended complex structure  that is easy to sample. These numerical results are supported by an analytical study of the training error landscape, and we show numerically that the qualitative features of the loss landscape are robust across different data structures. 
		Our study aims to provide new methodological insights for developing scalable methods for large networks.

	\end{abstract}

	\maketitle

	\section{Introduction}
	
	Understanding the geometry of the loss landscape in deep neural network models is a critical challenge~\cite{epfl_workshop}, as it directly influences the design and optimization of learning algorithms. The non-convexity of the loss landscape introduces significant complexity that often precludes the application of analytical methods.
	
	Empirical evidence suggests that first-order optimization methods such as gradient descent (GD) and its variants such as stochastic gradient descent (SGD) can effectively navigate certain regions of these landscapes~\cite{lecun2015deep,lecun1998gradient,bottou2010large,Kingma2014AdamAM,sgdvsadam}. However, the insights derived from such methods are closely tied to the specifics of the algorithms, limiting their usefulness in providing a comprehensive understanding of the underlying geometric structures. For example, large language models are typically not trained to optimality on their data~\cite{kaplan2020scaling,Hoffmann2022}, linking their generalization properties to high-loss configurations, a relationship that remains poorly understood but holds promise for improving learning efficiency.
	
	Analytical progress has been made in shallow, non-convex networks under simplified assumptions~\cite{gardner1988optimal,relu_locent}, using statistical physics techniques such as the replica and cavity methods. These studies reveal a highly intricate structure of minima~\cite{baldassi2020shaping,baldassi2015subdominant,baldassi2021unveiling,Annesi2023star}, characterized by features such as the overlap gap that renders some minima inaccessible~\cite{huang2014origin}, as well as rare, broad, and accessible minima. These minima have also been found to exhibit a diverse and broad range of generalization abilities~\cite{baldassi2022learning,baldassi2020wide}. 
	
	On the empirical side, simple low-dimensional visualizations of the loss landscape have been used to gain insight into the general properties of these minima and their arrangement within the loss landscape~\cite{LiVisualizing2018, Huang_Visualizations}. In particular, linear and piecewise linear paths~\cite{Draxler,garipov2018,fort2019large} have been used to show that weight configurations found by SGD are generally not linearly connected but are nevertheless connected through a piecewise path~\cite{pittorino2022deep,entezari2022the}.
	
	The problem of studying the geometry of the solutions of a deep neural network can be easily formulated in terms of statistical mechanics. While an exhaustive search for the regions associated with a low value of the loss $\mathcal{L}$ in the high-dimensional parameter space is  infeasible, one can explore the parameter space requiring the average loss $\langle\mathcal{L}\rangle$ to be small, while allowing its instantaneous value to fluctuate around it. 
	
	
	We investigate the loss landscape using a Hybrid Monte Carlo (HMC) approach \cite{Duane1987}, which enables large exploratory steps guided by gradients, in contrast to the purely random updates of standard Monte Carlo methods.
	
	
	We also introduce a Ratchet Hybrid Monte Carlo (RHMC) algorithm, which steers sampling along complex paths while maintaining low loss values. This approach is inspired by a similar technique shown to be efficient in protein folding simulations \cite{Camilloni2011} and in identifying their transition states \cite{Tiana2012}.

	To demonstrate the effectiveness of HMC and RHMC, we compare their results with analytical predictions for shallow non-convex networks and find agreement between the numerical results of the numerical methods and these predictions. Additionally, we uncover previously unreported theoretical insights into the geometry of minima in single--hidden--layer neural networks with generic activation functions~\cite{relu_locent}, including configurations that pose challenges to conventional optimization approaches.
	
	Although these results underscore the efficacy of HMC and RHMC as powerful tools for probing neural networks with intricate architectures, extending these findings to large-scale networks remains a challenge. Our work provides new insights in this regard.
	
	The architecture we have chosen to study the solution space of artificial neural networks is a tree committee machine ~\cite{barkai1992broken,engel1992storage,relu_locent}.  It can be considered as the simplest non-convex and non-linear toy model of a neural network. It has the advantages of being analytically treatable using replica methods techniques and that its optimized parameters are not related by trivial permutational symmetries. 
	Classical works in the 90s have studied this model by using the replica method~\cite{barkai1990statistical,engel1992storage} for the sign activation function and in the thermodynamic limit $N\to \infty$ and $P \to \infty$ with $\alpha \equiv P/N$ fixed and for $K=O(1)$. The typical and atypical states~\cite{relu_locent} of this model were studied in the large width $K$ limit (but with $K/N \to 0$) for a generic activation function. The same work provided a determination of the SAT/UNSAT transition, i.e. the maximum number of samples that the model can in principle store, in the Replica Symmetric (RS) and 1-step Replica Symmetry Breaking (1RSB) scheme. Recently the exact SAT/UNSAT transition has been computed in~\cite{annesi2024exact} by using a numerical solution of the full Replica Symmetry Breaking (fRSB) equations and it has been compared with the the maximal capacity reached by Gradient Descent; this unveiled the presence of an hard phase for Gradient Descent.

	A substantial body of work exists on the theoretical analysis of single-hidden-layer neural networks with fully connected receptive fields in the first layer, both when trained to store random patterns~\cite{engel1992storage, urbanczik1997full, nishiyama_solution_2025} or when the label is given by a teacher with a similar architecture~\cite{schwarze_statistical_1992, schwarze_learning_1993,  schwarze_discontinuous_1993,schwarze_learning_1993_1, aubin_committee_2018, oostwal_hidden_2021, citton_phase_2025,barbier2025generalization}. Recently, the scaling regime $K = O(N)$ has also been examined in~\cite{barbier2025statistical}.

	Our results are presented as follows. We shall first introduce the model and the algorithms to sample the loss space (Sect.~\ref{sec::sampling&connecting}), then we shall describe the numerical results obtained close to the interpolation threshold (Sect.~\ref{sec::underparametrized}) . In the case of the overparametrized regime (Sect.~\ref{sec::overparametrized}), we shall first present the numerical results of the sampling and then compare it with the analytical calculations.  Finally, we will discuss how these results change when realistic, correlated data are used as input (Sect.~\ref{sec::correlated}), and we will draw the overall conclusions (Sect.~\ref{sec::conclusions}).

	\section{The model and the Methods for sampling and connecting solutions}~\label{sec::sampling&connecting}
	
	\subsection{The model and main definitions}
	\label{sec::model}
	
\begin{figure}[h]
		\centerline{\includegraphics[width=1\linewidth]{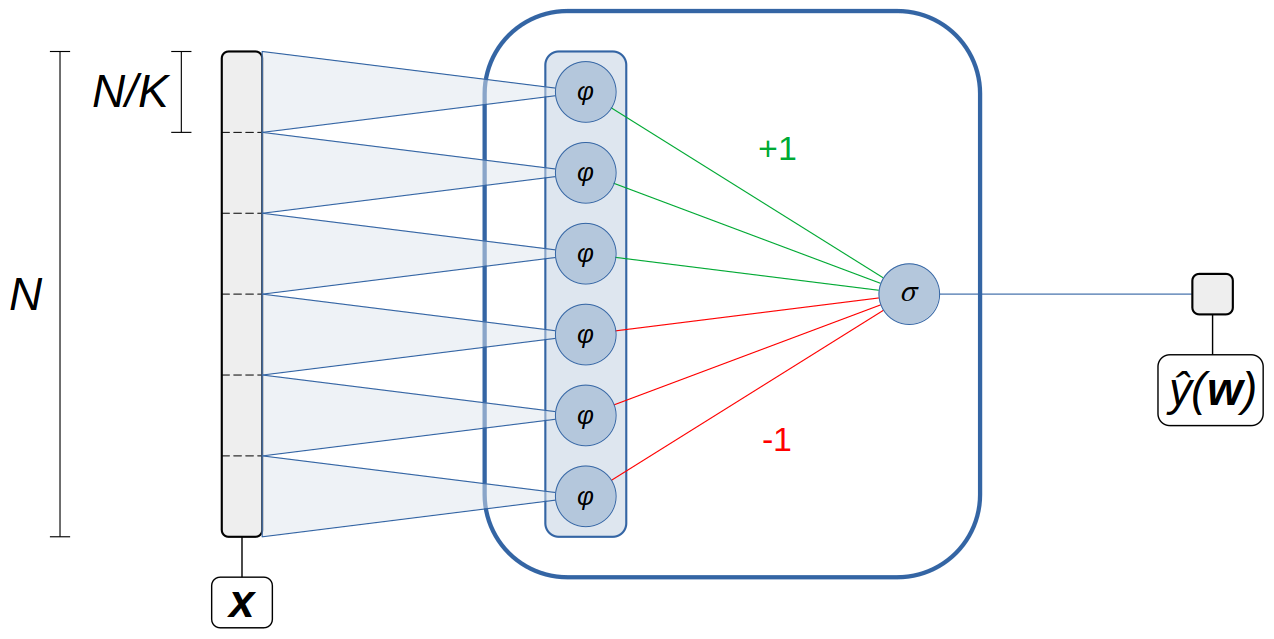}}
		\caption{Example of a tree committee machine with K=6, activation function $\varphi$ for the neurons in the hidden layer and $\sigma$ for the neuron in the output layer. The weights of the output neuron have been set to $\pm1$ for $i=1,...,\frac{K}{2}$ and $i=\frac{K}{2}+1,...,K$ respectively.}
		\label{fig:CommitteeMachine}
	\end{figure}

	The tree committee machine we consider consists of a two layer neural network with a generic non-linear activation function, in which each of the $K$ neurons of the hidden layer is connected only to a subset of $N/K$ elements of the $N$--dimensional input vector (Fig.~\ref{fig:CommitteeMachine}).  
	
	For any input $\boldsymbol{x}$, the output of the tree committee machine can be written as
	\begin{multline}
		\hat y(\boldsymbol{w}) = \text{sign} [ \Delta(\boldsymbol{w}) ]\\
		= \text{sign}\left[ \frac{1}{\sqrt{K}} \sum_{l=1}^K c_l \, \varphi\left( \sqrt{\frac{K}{N}} \sum_{i=1}^{N/K} w_{li} x_{li} \right) \right],
		\label{eq::tcm}
	\end{multline}
	where $K$ is the width of the hidden layer and $\varphi$ is a non-linear activation function. The first layer is parameterized by a set of weights $\w \in \mathbb{R}^{N}$, which will be trained to learn a dataset.
	The weights of the second layer, $c_l$ are fixed to $\pm 1$, with equal probability. Thus, the number of learned weights is exactly equal to $N$, the input dimension of the network. The main reason for not learning the weights of the second layer is to eliminate the symmetries of node permutations, which would make analyzing the space of solutions more difficult.
	
	We consider a synthetic dataset $\mathcal{D}$ made of $P$ input vectors $\boldsymbol{x}^\mu$, whose elements are extracted from a standard normal distribution, and the associated labels $y^\mu = \pm 1$, also generated randomly with equal probability. The ratio $\alpha=P/N$ is called \emph{constrained density}.
	
	The task is to find an $N$-dimensional vector $\boldsymbol{w}$ that correctly classifies each input $\boldsymbol{x}^\mu$ in the dataset to the corresponding label $y^\mu$, for every $\mu \in [P]$, i.e.
	\begin{equation}
		\label{eq::constraints}
		y^\mu \Delta^\mu(\boldsymbol{w}) > 0\,, \qquad \forall \mu \in [P] \,.
	\end{equation}
	A weight vector $\boldsymbol{w}$ satisfying Eq.~\eqref{eq::constraints} will be therefore called in the following as a \emph{solution} of the classification problem. 
	Equivalently, this means that the training error of $\boldsymbol{w}$ defined as the number of misclassified input-output associations
	\begin{equation}
		\label{eq::training_error}
		\begin{split}
			&\epsilon_t = \sum_{\mu = 1}^P \ell_{\mathrm{NE}}\left( y^\mu \Delta^\mu(\boldsymbol{w}) \right) \\
			&\ell_{\mathrm{NE}}\left( y^\mu \Delta^\mu(\boldsymbol{w}) \right) \equiv \Theta(- y^\mu \Delta(\boldsymbol{w}))
		\end{split}
	\end{equation} 
	is equal to zero, where we have denoted by $\Theta(\cdot)$ the Heaviside function, and by $\ell_{\mathrm{NE}}$ the ``error counting'' (or number of error) loss. Since this loss is not differentiable, it is not used for optimization in machine learning, which typically relies on gradient-based algorithms. The loss that is used to find solutions is usually called ``\emph{training loss}'' in the machine learning jargon. Instead, the error counting loss usually serves to evaluate whether a solution has been found\footnote{In general the training error $\epsilon_t$ could be zero even if the training loss is not.}, rather than guiding the optimization process itself. 
	
	A loss function that is commonly used to actually find solutions, and on which we will focus on this paper, is the so called \emph{cross-entropy} loss, which in binary classification reads
	\begin{equation}
		\label{eq::cross_entropy}
		\ell_{\mathrm{CE}}(y^\mu \Delta^\mu(\boldsymbol{w})) = \ln\left( 1 + e^{-y^\mu \Delta^\mu(\boldsymbol{w})} \right).
	\end{equation}  
	In the following we will say that a solution is ``\emph{typical}'' if it is extracted from the flat measure over the set of all solutions. Sometimes one requires not only that $\boldsymbol{w}$ is a solution of the classification problem, but also that it satisfies a certain degree of robustness.
	This can be enforced by ensuring that $\Delta^\mu(\boldsymbol{w})$ aligns with the corresponding label $y^\mu$ for any $\mu \in [P]$, within a specified confidence level $\kappa$
	\begin{equation}
		\label{eq::constraints_margin}
		y^\mu \Delta^\mu(\boldsymbol{w}) > \kappa \,, \qquad \forall \mu \in [P]
	\end{equation}
	$\kappa$ is also named ``\emph{margin}''. Imposing a margin $\kappa>0$ ensures that the solution sampled non only has zero training error, but also it is robust to perturbations of the inputs. Both typical (i.e. $\kappa = 0$) and atypically robust solutions with a positive margin $\kappa$ can be obtained using a loss function that generalizes the one defined in Eq.~\eqref{eq::training_error},
	\begin{equation}
		\label{eq::loss_margin}
		\ell(y^\mu\Delta^\mu(\boldsymbol{w})) = \Theta(-y^\mu \Delta^\mu(\boldsymbol{w}) - \kappa)	 \,.
	\end{equation}
	We stress that extracting solutions by optimizing a loss different from the error counting loss is therefore to be considered as ``atypical''.

	\subsection{The canonical--ensemble framework}
	We shall look for the solutions of the network using the formalism of the canonical ensemble, thus sampling the space of parameters that have in average a given value of the loss. The solutions of the network are then found sampling the parameters at low temperature, where the average loss is minimal.
	
	The posterior distribution corresponding to a given loss function (or log-likelihood) $\mathcal{L}(\boldsymbol{w}; \mathcal{D})$ is given by the Boltzmann-like distribution
	\begin{equation}
		\label{eq::Boltzmann}
		p(\boldsymbol{w}, \beta; \mathcal{D}) = \frac{e^{-\beta \mathcal{L}(\w; \mathcal{D})} p(\w)}{Z(\beta; \mathcal{D})}
	\end{equation}
	where $\beta=1/T$ is the inverse temperature, $p(\w)$ is the prior distribution of the weights. For simplicity, a Boltzmann constant $k_B = 1$ is assumed in every equation. The factor $Z(\beta; \mathcal{D})$ is the partition function and it is equal (up to a normalization factor) to what is called \emph{evidence} in Bayesian statistics; it reads
	\begin{equation}
		\label{eq::partition_function}
		Z(\beta; \mathcal{D}) = \int d\w \, p(\w) \, e^{-\beta \mathcal{L}(\w; \mathcal{D})}
	\end{equation}
	The loss function usually considered in the machine learning literature is separable into a sum over the $P$ elements of the dataset, i.e.
	\begin{equation}
		\mathcal{L}(\boldsymbol{w}; \mathcal{D}) \equiv \sum_{\mu = 1}^{P} \ell(y^\mu \Delta^\mu(\boldsymbol{w}))
	\end{equation}
	where $\ell$ is a loss function per pattern and $\Delta(\boldsymbol{w})$ identifies the preactivation of the output node. A solution satisfying Eq.~\eqref{eq::constraints_margin} can be obtained from the canonical-ensemble framework by performing the limit $\beta \to \infty$ of the Boltzmann measure in Eq.~\eqref{eq::Boltzmann}, equipped with the loss function of Eq.~\eqref{eq::loss_margin}.
	
	In the following we consider as a prior a standard normal distribution, or equivalently a $L^2$ regularization term with parameter $\lambda$
	\begin{equation}
		\label{eq::prior}
		p(\w) = e^{-\frac{\beta \lambda}{2} |\w|^{2}}
	\end{equation}
	We can incorporate the regularization term and the loss term in a function $U(\boldsymbol{w})$
	\begin{equation}
		U(\w) = \mathcal{L}(\w; \mathcal{D}) + \frac{\lambda}{2} |\w|^{2}
		\label{eq:potentialfunc_impl}
	\end{equation}
	that can be though as the (potential) energy associated to the neural network parameters $\w$. Equation~\eqref{eq::Boltzmann} can be then rewritten as
	\begin{equation}
		p(\w, T) = \frac{e^{-\frac{U(\w)}{T}}}{Z_{U}(T)}
		\label{eq:p(theta)}
	\end{equation}
	where we have dropped for simplicity the dependence on the dataset $\mathcal{D}$ both on the posterior distribution and in the partition function $Z_{U}(T) = \int d\w \, e^{-\frac{U(\w)}{T}}$.

	\subsection{Hybrid Monte Carlo algorithm}
	\label{sec:methods_HMC}
	
	One of the main goals of statistical inference is to be able to sample from the posterior distribution of Eq.~\eqref{eq:p(theta)}. Monte Carlo--based methods achieve this by simulating a Markovian stochastic process over $\w$ that converges to the distribution of Eq.~(\ref{eq:p(theta)}). 
	In the Metropolis implementation, the transition rate of the stochastic process is
	\begin{multline}
		r_{M}(\boldsymbol{w}_i \rightarrow \boldsymbol{w}_{i+1}) \\
		= r_0 \, p_{ap}(\boldsymbol{w}_{i+1}|\boldsymbol{w}_i) \, \mathrm{min}\biggl[ 1, e^{-\frac{U(\boldsymbol{w}_{i+1}) - U(\boldsymbol{w}_i)}{T}} \biggr]
		\label{eq:wM}
	\end{multline}
	where $r_0$ is a rate constant, $p_{ap}(\boldsymbol{w}_{i+1}|\boldsymbol{w}_i)$ is the {\it a priori} conditional probability of proposing a move and  the last term is the acceptance probability. If $p_{ap}(\boldsymbol{w}_{i+1}|\boldsymbol{w}_i)$ is symmetric upon exhange of the two states, the condition of detailed balance holds and the process converges to Eq. (\ref{eq:p(theta)}). Often a uniform {\it a priori} probability is used; however, the higher the dimension of the system, the lower is the probability of going in an optimal direction with a uniform choice.
	
	A way of mitigating this problem is to choose the {\it a priori} probability exploiting the knowledge of the energy  gradient \cite{Duane1987}, using a Hamiltonian formalism. Defining $\p \in \mathbb{R}^{N}$ as the momenta associated to the weights $\w$, the Hamiltonian of the system is 
	\begin{equation}
		\begin{split}
			H(\w, \p) &= \frac{1}{2} \p^{T} \mathcal{M}^{-1} \p + U(\w) 
		\end{split}
		\label{eq:Hamiltonian}
	\end{equation}
	where $\mathcal{M}^{-1}$ is the inverse of a (fictitious) mass matrix. 
	
	The solutions to the equations of motion resulting from this Hamiltonian can be used as a preliminary step in the HMC algorithm. The time reversal invariance of Hamiltonian systems guarantees that the detailed balance holds.
	
	In general, the equations of motions are solved with a numerical integrator, and detailed balance is satisfied only in the limit of time step $\delta t \rightarrow 0$. However, some integrators like the velocity Verlet algorithm
	\begin{equation}
		\begin{split}
			\w^{(k+1)} &= \w^{(k)} + \mathcal{M}^{-1}\p^{(k)}\delta t - \mathcal{M}^{-1}\nabla U(\w^{(k)})\frac{\delta t^2}{2} \\
			\p^{(k+1)} &= \p^{(k)} - \frac{\delta t}{2} \bigl[ \nabla U(\w^{(k+1)}) + \nabla U(\w^{(k)}) \bigr].    
		\end{split}
		\label{eq:VelVerlet}
	\end{equation}
	are intrinsically symmetric for time reversal, and thus detailed balance holds for any choice of $\delta t$, even if the energy is not conserved. In fact, evaluating the latter of Eqs.~\eqref{eq:VelVerlet} for 
	$\delta t \to -\delta t$ one obtains 
	\begin{align}
		\p^{(k+1)} + \frac{\delta t}{2} \nabla U(\w^{(k+1)}) = \p^{(k)} - \frac{\delta t}{2} \nabla U(\w^{(k)}),
	\end{align}
	and substituting it in the former, 
	\begin{align*}
		\w^{(k)} &= \w^{(k+1)} - \mathcal{M}^{-1}\p^{(k+1)}\delta t - \mathcal{M}^{-1}\nabla U(\w^{(k+1)})\frac{\delta t^2}{2} \\
		\p^{(k)} &= \p^{(k+1)} + \frac{\delta t}{2} \bigl[ \nabla U(\w^{(k)}) + \nabla U(\w^{(k+1)}) \bigr],
	\end{align*}
	which is a backward trajectory with respect to that of Eqs.~\eqref{eq:VelVerlet}. 
	
	Operatively, at each iteration of the Metropolis algorithm, the momenta are extracted from a Maxwell--Boltzmann distribution at temperature $T$ and a trajectory in the phase space is generated solving Eqs. (\ref{eq:VelVerlet}). The final point of the trajectory is then accepted with probability $\min(1,\exp[-\Delta H/T])$, where $\Delta H=H(\boldsymbol{w}_{i+1},\boldsymbol{p}_{i+1})-H(\boldsymbol{w}_{i},\boldsymbol{p}_{i})$. In the condition of detailed balance (i.e. when a time--reversible integrator is employed), the kinetic energy arising from the Metropolis acceptance probability and that coming from the Maxwell--Boltzmann term associated with the extraction of the random momenta cancel out. This leads the distribution of the states sampled by HMC to converge to Eq.~\eqref{eq:p(theta)}.
	
	The power of this scheme is that it can produce large moves in a direction that smoothly changes the energy of the system, and thus the Metropolis acceptance rate can remain high.
	
	\subsection{Double ratchet}
	\label{sec:methods_DoubleRatchet}
	A key line of research in characterizing the structure of neural network loss landscapes explores pathways between distinct weight configurations $\boldsymbol{w}$ and $\boldsymbol{w}'$, which typically correspond to solutions with low loss~\cite{goodfellow2014qualitatively,garipov2018,fort2019large,li2018visualizing,frankle2020revisiting}. The simplest approach examines how the loss behaves along the linear (or straight) path connecting these points. As shown in~\cite{Draxler}, overparameterized neural networks trained with stochastic gradient descent (SGD) often exhibit a loss barrier along the linear path. The barrier can be notably decreased by removing the symmetries that the neural network possess, like permutation symmetry of the hidden units~\cite{pittorino2022deep,entezari2022the}.
	
	While linear paths provide an intuitive and computationally efficient visualization of the loss landscape, they do not explicitly search for paths between solutions. To address these limitations, the authors of~\cite{garipov2018} introduced the use of piecewise linear trajectories, or ``polygonal chains'', where the path is optimized by introducing $k$ intermediate pivot points. Although this method allows for more flexible connections, it is computationally demanding, requiring multiple training runs to optimize the pivot locations. Furthermore, the number of required pivots can grow significantly, particularly in less overparameterized settings, making the approach increasingly impractical in such regimes.
	
	As an alternative to generate trajectories from $\w$ to $\w'$, we make use of a modified HMC algorithm that damps fluctuations in the direction opposite to $\w'$. This is inspired from the ratchet--and--paw mechanical system and has given good results in molecular dynamics simulations \cite{Camilloni2011}. It can be implemented calculating the dynamics of two points $\w_1(t)$ and $\w_2(t)$ starting in $\w$ and $\w'$, respectively, and evolving under a HMC with energy
	\begin{multline}
		U_{rt} \bigl[ \boldsymbol{w}_1, \boldsymbol{w}_2, t \bigr]
		= U \bigl[ \boldsymbol{w}_1 \bigr] + U \bigl[ \boldsymbol{w}_2 \bigr] + \widetilde{U} \bigl[ \boldsymbol{w}_1, \boldsymbol{w}_2, t \bigr]
		\label{eq:ratchetfunc_impl}
	\end{multline}
	where $U$ is the potential energy of Eq.(\ref{eq:potentialfunc_impl}) and
	\begin{widetext}
		\begin{equation}
			\widetilde{U} \bigl[ \boldsymbol{w}_1(t), \boldsymbol{w}_2(t), t \bigr] = 
			\begin{cases}
				\frac{k}{2} \bigl[ |\boldsymbol{w}_1(t)-\boldsymbol{w}_2(t)| - d_{\min}(t) \bigr]^2 & \text{if } |\boldsymbol{w}_1(t)-\boldsymbol{w}_2(t)| > d_{\mathrm{min}}(t) \\
				0 &  \text{if } |\boldsymbol{w}_1(t)-\boldsymbol{w}_2(t)| \leq d_{\min}(t)
			\end{cases}
			\label{eq:ratchetfunc_expl}
		\end{equation}
	\end{widetext}
	where $d_{\mathrm{min}}(t) \equiv \min_{t'<t} |\boldsymbol{w}_1(t')-\boldsymbol{w}_2(t')|$ is the minimum distance observed along the trajectories of the two points up to time $t$. The time--dependent term of Eq.~(\ref{eq:ratchetfunc_expl}) favors the moves which get the two weights closer to each other, without exerting work to push them. In this way, the two points can autonomously find the minimum--energy path that connects them. 
	
	A suitable range of values for the harmonic constant k can easily be estimated by trial and error. At high k, one of the two vectors becomes trapped along suboptimal paths, resulting in a lower acceptance rate as the system becomes stuck between the steep parabolic potential and the energy barrier separating it from the other vector. In these cases, reducing the steepness of the parabolic potential allows for larger oscillations in the distance between the two weights.
	
	In practice, each iteration consists of two sequential updates, one for each vector. During each update, the HMC scheme uses the potential $U_{rt} \bigl[ \boldsymbol{w}_1(t), \boldsymbol{w}_2(t) \bigr]$  is used for the selected vector, while the other vector is kept fixed. This process is repeated until the cosine similarity (or normalized overlap) between the two weights
	\begin{equation}
		q(\boldsymbol{w}_1, \boldsymbol{w}_2) = \frac{\w_1 \cdot \w_2}{|\w_1| |\w_2|} 
		\label{eq:qcos}
	\end{equation}
	is $>0.995$ (i.e., $q\approx 1$). We will call the point $\s^\star$ where the two trajectories $\boldsymbol{w}_1(t)$ and $\boldsymbol{w}_2(t)$ meet as the ``\emph{anchor}'' weight of $\boldsymbol{w}_1$, $\boldsymbol{w}_2$.   
	
	
	\subsection{Coupled replica simulations}
	\label{sec:methods_Replica}
	
	Besides sampling the Boltzmann--like probability of Eq.~\eqref{eq:p(theta)}, it can be useful to bias in order to give more statistical weight to wider energy minima, which were shown to have better generalization properties than narrow ones~\cite{baldassi2015subdominant,baldassi2020wide,baldassi2021unveiling,baldassi2022learning}. 
	
	We did this using an entropy--driven search algorithm inspired from ref.~\cite{unreasoanable}. We used a system composed of $y$ coupled replicas $\{\boldsymbol{w}_i(t)\}_{i=1}^{y}$, each starting from a different solution previously found by using the HMC. The replicas are coupled to the barycenter of the system
	\begin{align}
		\w_{c}(t) = \frac{1}{y} \, \frac{\sum_{i=1}^{y} |\boldsymbol{w}_i(t)|}{|\sum_{i=1}^{y} \boldsymbol{w}_i(t)|} \, \sum_{i=1}^{y} \boldsymbol{w}_i(t)
		\label{eq:bary}
	\end{align}
	with a potential 
	\begin{equation}
		U_{rp}(\{\boldsymbol{w}_i(t)\}_{i=1}^{y}) = \sum_{i=1}^{y} U \bigl[ \boldsymbol{w}_i(t) \bigr] + \frac{\gamma T}{y} \sum_{i=1}^{y} |\boldsymbol{w}_i(t)-\w_{c}(t)|,
		\label{eq:replicafunc}
	\end{equation}
	where $\gamma$ is a Lagrange multiplier regulating the mean distance between the replicas and the barycenter $\w_{c}$.
	
	The implementation of the replica simulations is similar to that of the double-ratchet. Each iteration consists of a number of sequential updates, equal to the number of replicas. During each update, the HMC scheme with potential $U_{rp}(\{\boldsymbol{w}_i(t)\}_{i=1}^{y})$ is adopted for the selected replica, while the other ones are kept fixed. The barycenter vector is also updated, both during the proposed move and at the end of the acceptance procedure, for each replica. 
	The value of $\gamma$ is increased during the simulations, until the $y$ replicas $\boldsymbol{w}_i$ all collapse onto a single high--entropy weight $\boldsymbol{c}$, that we will refer in the rest of the paper as the ``\emph{center}'' weight found by the coupled replica simulation.
	
	Note that in the definition of the barycenter of Eq.~\eqref{eq:bary}, the norm is rescaled to match the mean of the other replicas. This prevents the replicas from being driven toward a lower norm point, which would undesirably increase their energy $U \bigl[ \boldsymbol{w}_i(t) \bigr]$.

	\section{Results}
	
	Using the numerical techniques listed in the previous section, we explored the learning loss landscape of the tree committee machine at different energy levels and in different regimes of constrained density.  
	Our main results are the following:
	
	\begin{itemize}
		\item Close to interpolation threshold, the low-energy manifold has a spiky shaped structure, i.e. it is composed by sharp protrusions (targeted by GD) from which HMC is not able to escape at low temperatures. We nevertheless show that those protrusions are connected by complex paths through a more compact narrow region that is not completely flat (see section~\ref{sec::underparametrized})
		\item In the overparametrized regime $\alpha \ll 1$ we show that the spikes do not trap anymore HMC. Furthermore, the low-energy manifold is entirely flat in the bulk, giving it a star-shaped structure.
		We confirm this numerical evidence by studying analytically the error landscape, showing analogous results (see section~\ref{sec::overparametrized}).
		\item Preliminary experiments on highly correlated, real-world datasets indicate that our findings remain robust even when the data exhibit significant structural properties. 
	\end{itemize}
	
	Unless stated otherwise, in all the numerical simulations we have used a tree committee machine with $N=1000$ input neurons, $K=50$ hidden neurons, and ReLU activation functions $\varphi(x) = \max(0, x)$. In HMC simulations we have used the cross entropy loss function in Eq.~\eqref{eq::cross_entropy}. The norm of the weights is controlled by the Lagrange multiplier $\lambda=2P \cdot 10^{-7}$, with $P=\alpha N$ being the dimension of the dataset. The mass matrix in Eq.~\eqref{eq:Hamiltonian} required by the algorithm is set $\mathcal{M} = \mathbb{I}$.
	
	\begin{figure*}
		\centerline{\includegraphics[width=\linewidth]{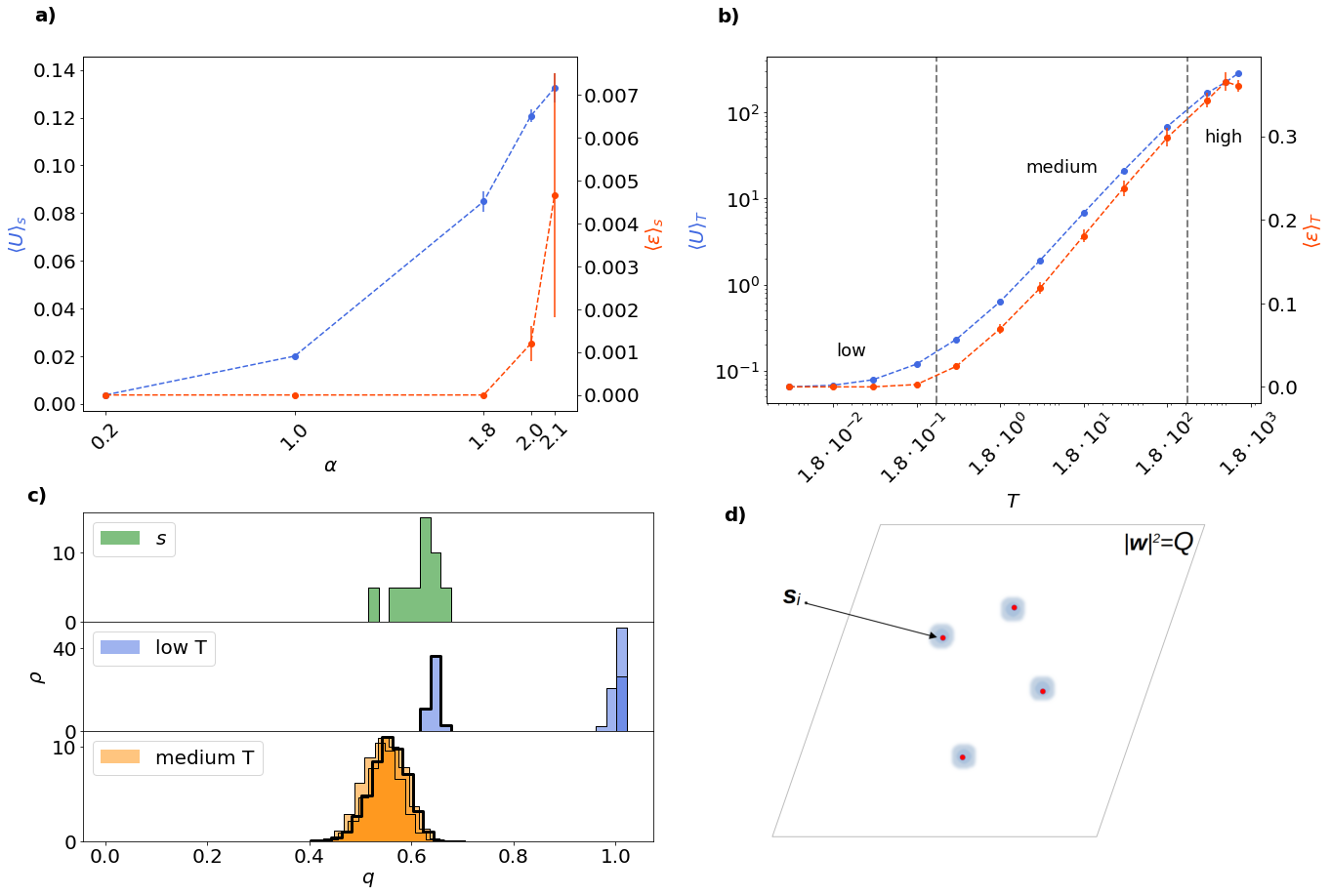}}
		\caption{\textbf{a)} The average energy $\langle U\rangle_{\s}$ (blue dots) and the mean training error $\langle\epsilon\rangle_{\s}$ (orange dots) computed on GD solutions with respect to the constraint density $\alpha = P/N$.
			\textbf{b)} The average energy $\langle U\rangle_{T}$ (blue dots) and the mean training error $\langle\epsilon\rangle_{T}$ (orange dots) with respect to the temperature $T$. The gray vertical lines identify the three temperatures regimes (low, medium and high). 
			\textbf{c)} The upper panel shows the cosine--similarity distribution among GD solutions $\s$ at $\alpha=1.8$ (in green). The thin--line histogram in the middle panel (in blue) shows the two intra-state overlap distributions $\rho(q)$ in the  low--temperature ($T=1.8\cdot10^{-2}$) simulations starting from two different initial conditions. The thick line histogram stands for the corresponding inter-state distribution computed between the two sets of samples states.
			Two thin--line histograms in the lower panel (in orange) shows intra-state  overlap distributions $\rho(q)$ relative to two low--temperature ($T=1.8\cdot10^{1}$) simulations starting from different initial conditions. The thick line histogram stands for the corresponding inter-state distribution computed between the two sets of samples states.
			In the lower panel, the three overlapping distributions have been slightly shifted along the $x$--axis to facilitate the comparison.
			\textbf{d)} A sketch of the subspace of solutions at fixed norm as suggested by the HMC simulations.}
		\label{fig:FixedTempMonteCarlo}
	\end{figure*}

	\subsection{The model close to the interpolation threshold} \label{sec::underparametrized}
	
	We first study the loss landscape of the model in the underparametrized regime, i.e. just below the threshold at which full--batch GD can no longer find weights with zero training error and which is usually called "interpolation threshold"~\cite{engel-vandenbroek,Belkin2019}. This happens approximately around $\alpha = 1.8$ (Fig.~\ref{fig:FixedTempMonteCarlo}a), which is very far from the SAT/UNSAT transition $\alpha_c \simeq 2.65$ where solutions cease to exist and which was computed in~\cite{annesi2024exact}. All the numerical studies of this section therefore refer to the case $\alpha=1.8$ (i.e. $P=1800$). The results below are obtained from different simulations, with the same dataset, which have reached a stationary regime. Given the dimension $N$ of the network and the dataset size $P$, we believe that these results are not instance-dependent.
	
	\subsubsection{The weight space displays three regimes with respect to the temperature} \label{sec::results_HMC}
	
	First, we trained the model with GD for approximately $\sim10^{7}$ epochs and with a fixed learning rate $\eta = 1.0$. The solutions $\s_{i}$ found with GD, from which we started the sampling, have mean energy $\langle U\rangle_{\s} = (8.5 \pm 0.04)\cdot10^{-2}$ and mean training error $\langle\epsilon\rangle_{\s}=0$ (Fig.~\ref{fig:FixedTempMonteCarlo}a).
	The similarity between them is peaked at $q \approx 0.6$ (see upper panel in Fig.~\ref{fig:FixedTempMonteCarlo}c), which means that GD finds solutions with an almost fixed mutual distance.
	
	Starting from GD solutions, we explored the space of parameters of the network with the HMC algorithm at different temperatures (Fig. \ref{fig:FixedTempMonteCarlo}b). For $T<1.8\cdot10^{-1}$ the average energy $\langle U\rangle_{T}$ starts to flatten, suggesting that here the system freezes in the lowest--energy available states. 
	The average energy is comparable with that of the solutions $\s_{i}$ and also the mean training error remains $\langle\epsilon\rangle=0$. At these temperatures the sampling of the space of weights is difficult and the system gets trapped in local minima of the loss. The similarity distribution $\rho(q)$ (see Eq.~(\ref{eq:qcos})) calculated on weights sampled from HMC trajectories starting from the same initial condition (\emph{intra-state} overlap distribution) is strongly peaked around 1, see Fig.~\ref{fig:FixedTempMonteCarlo}c, middle panel. This is markedly different from the \emph{inter-state} overlap distribution, which is obtained by measuring the overlap between weights along HMC trajectories starting from strictly different initial conditions and is peaked at $q \approx 0.63$. Thus, the system is not able to equilibrate at these temperatures but stays close to the initial condition.
	
	In the range of temperature $1.8\cdot10^{-1}<T<1.8\cdot10^{2}$ the average energy increases almost linearly $\langle U \rangle_{T} \sim T^{\delta}$, with a scaling exponent $\delta \approx 0.989 \pm 0.019$. This is typical of a glassy thermodynamic phase with a sub-exponential number of accessible states.
	In this intermediate temperature range, there is no marked difference between the inter and intra state overlap distribution (Fig.~\ref{fig:FixedTempMonteCarlo}c, lower panel). This suggests that the system is not trapped in local minima as in the low--temperature phase. The average similarity between states displays a single peak centered at $q\approx 0.6$, similar to that observed at low temperature, but with a larger variance.
	
	In the high temperature phase, the average energy deviates from the linear behavior and the training error further increases. The distribution of similarities $q$ is very broad. However, here the HMC algorithm becomes inefficient because the high velocities extracted before each HMC move require a very small step $\delta t$ for a correct integration of the equations of motion and thus for an acceptable acceptance rate. 
	
	We summarize these results in Fig.~\ref{fig:FixedTempMonteCarlo}d, where we show a sketch of the space of parameters at fixed norm. On this hypersphere, the solutions found by the GD define energy basins which are explored by the system at low and intermediate temperatures. 
	These basins are disconnected at low temperature, since HMC remains confined near the initialization and cannot find pathways connecting the different basins while remaining at low energy and at zero training error.

	\begin{figure*}[t]
		\centerline{\includegraphics[width=\linewidth]{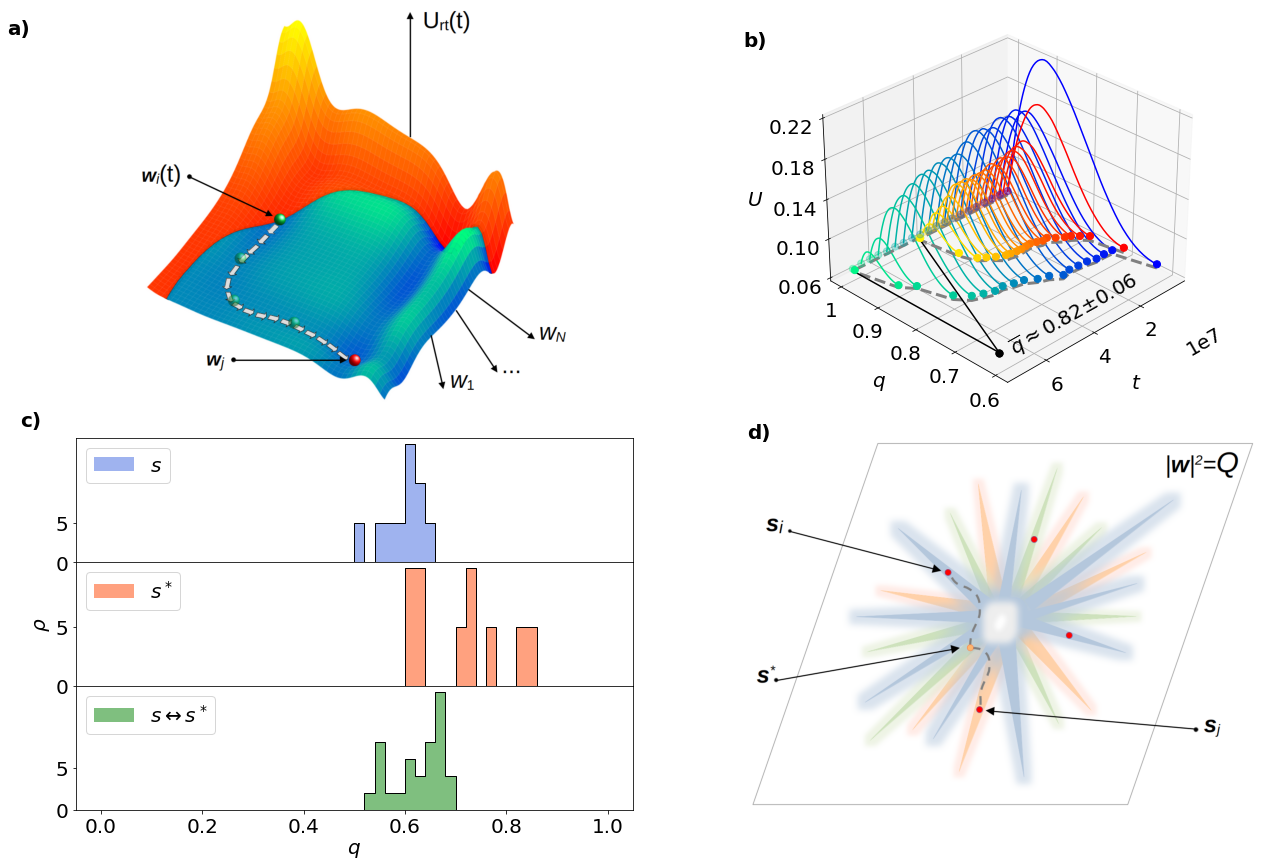}}
		\caption{\textbf{a)} A sketch of the ratchet algorithm. The point $\boldsymbol{w}_i(t)$  (green sphere) reaches the (fixed) point  $\boldsymbol{w}_j$ (red sphere), through low energy barriers. Backward moves (red region) are dumped by the quadratic term of~\eqref{eq:ratchetfunc_expl}.
			\textbf{b)} The energy profile along the geodesic evolving during the simulation time $t$ with respect to the similarity $q$ between the two moving points of the double ratchet (dashed line). Two different ratcheted trajectories are shown starting from the same weights at $T=T_L\equiv1.8\cdot10^{-2}$.
			The solid curves indicate the energies calculated along geodesics connecting pairs of points picked at the same time. The mean and the standard deviation of the similarity between the anchor weights $\s^{*}$ obtained from independent trajectories are also indicated. 
			\textbf{c)} The similarity distribution $\rho(q)$ between the GD solutions $\s$ (in blue, upper panel), between the double--ratchet anchor weights $\s^{*}$ (in orange, middle panel) and between $\s$ and $\s^{*}$ (in green, bottom panel). 
			\textbf{d)} A sketch of the low--energy manifold.}
		\label{fig:DoubleRatchet}
	\end{figure*}
	
	\subsubsection{The low--energy basins are connected by complex paths}
	\label{sec:results_DoubleRatchet}

	To explore the connectivity of energy basins at low temperatures, we performed double-ratchet simulations, hoping to find paths which are not found by the HMC algorithm. This method is designed to identify low-barrier pathways between two weight configurations by following the minimal gradient in the direction that brings them closer together  (Fig.~\ref{fig:DoubleRatchet}a).
	
	The double ratchets are initialized on different pairs of solutions $(\s_i, \s_j)$ found after a short thermalization of the system of approximately $10^4$ HMC steps. Along each simulation, the weights do not cross significant energy barriers, as compared to the average energy at this temperature, keeping the training error equal to zero. Conversely, the \emph{geodesic} between pairs of points (i.e. the linear interpolation between weights at the fixed value of the norm imposed by the two endpoints) along the double--ratchet trajectory results in barriers that are substantially higher than the average energy (Fig. \ref{fig:DoubleRatchet}b). 
	These results suggest that, although the solutions found by gradient descent are linearly mode disconnected, low-energy tortuous paths joining them still exist.
	
	The anchor point $\s^\star$ obtained at the end of the double--ratchet trajectory has an energy that is slightly lower than the one of the initial points $(\s_i, \s_j)$ and $\langle U\rangle_{\s^\star} = (6.80 \pm 0.09) \cdot10^{-2}$ is compatible with the average potential energy at $T_L=1.8\cdot10^{-2}$, that is the constrained dynamic arising from the double--ratchet simulation helps the equilibration of the system. The anchor weights $\s^\star$ obtained from several double ratchet simulations initialized from different weight pairs $(\s_i, \s_j)$ exhibit larger similarity than the respective initial vectors and are separated by energy barriers along their linear interpolation (Fig. \ref{fig:DoubleRatchet}c). This suggests the presence of a compact and narrow region of the solution space that must be traversed to move from one gradient descent solution to another. The similarity between the anchor weights $\s^\star$ and the associated initial weights is also centered around $0.6$, as that between the starting solutions.
	
	In summary (Fig.~\ref{fig:DoubleRatchet}d), the basins identified by GD solutions are connected by non--linear paths pointing towards a denser, more compact region with slightly lower energy. The low--energy manifold of the space of weights has thus a spiky shape, with the GD solutions lying on its spikes.

	\subsubsection{The center is not flat}
	\label{sec:results_Replica}
	\begin{figure*}
		\centerline{\includegraphics[width=\linewidth]{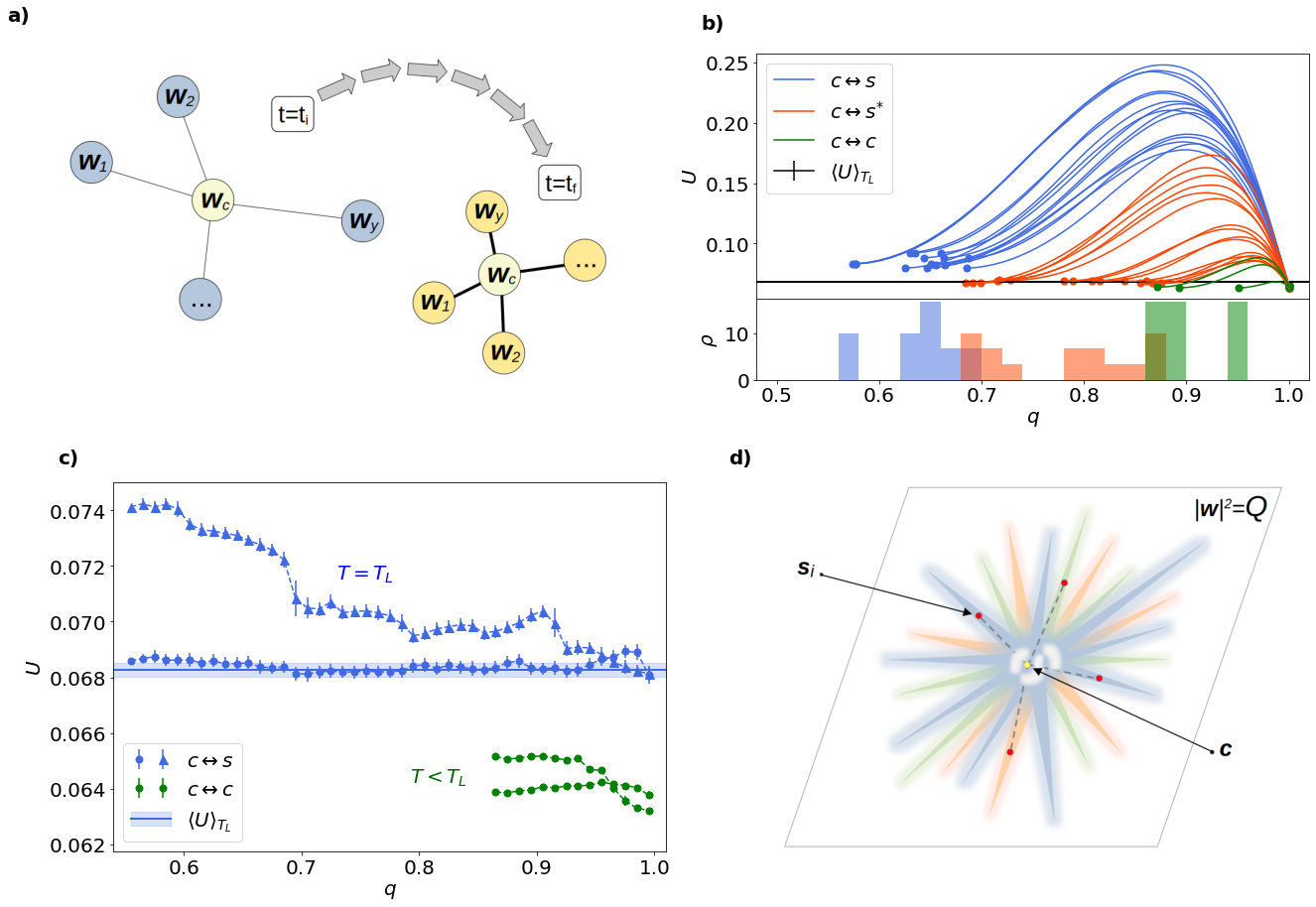}}
		\caption{\textbf{a)} A sketch of the replica algorithm. The $\{\boldsymbol{w}_i\}_{i=1}^{y}$ points are coupled to their barycenter $\w_c$ and the coupling constant is slowly increased during the simulation, until all the replicas collapse onto one small region of the weights space.  
			\textbf{b)} The energy profile along the geodesic at fixed norm between centers and GD solutions ($\boldsymbol{c}\leftrightarrow\boldsymbol{s}$), between centers and double--ratchet solutions ($\boldsymbol{c}\leftrightarrow\boldsymbol{s}^*$) and between centers ($\boldsymbol{c}\leftrightarrow\boldsymbol{c}$) (upper panel). 
			The black line marks the mean energy (and standard deviation) at $T=T_L$. In the lower panel, the similarity distribution $\rho(q)$ between centers and GD solutions (blue), between centers and double--ratchet solutions (orange) and among centers (green). 
			\textbf{c)} The potential energy $U$ along the trajectory obtained by performing double--ratchet simulations connecting a center vector to a GD solution at $T=T_L$ (lower and upper blue curve, respectively) and two center points at $T=5.4\cdot10^{-3}<T_L$ (green curves), where the similarity $q$ parametrizes the trajectory.
			The blue line and shaded area are the mean energy and standard deviation, respectively, at $T=T_L$. 
			\textbf{d)} A sketch of the subspace of solutions where the central region has been studied by coupled replica simulations.}
		\label{fig:Replica}
	\end{figure*}	    
	
	We then studied the geometry of the center of the spiky low energy manifold in more detail using coupled replica simulations, starting with five GD solutions coupled to their barycenter and increasing the coupling term $\gamma$ (Eq.~\eqref{eq:replicafunc}) until all the running points converge to a single vector $\boldsymbol{c}$ (Fig.~\ref{fig:Replica}a). 
	
	We compared the properties of the centers found in different coupled replica simulations with the solutions $\s$ found by GD and the anchor points $\s^{\star}$ found by double ratchets initialized on different GD solutions at $T=T_L$. The final configuration $\boldsymbol{c}$ always has an energy that is lower by several standard deviations than the mean energy at the simulation temperature (Fig.~\ref{fig:Replica}b,c). 
	The centers obtained from the coupled replica simulations are confined to a very narrow region of the solution space (Fig.~\ref{fig:Replica}b), even narrower than the region spanned by the anchor points (middle panel of Fig.~\ref{fig:DoubleRatchet}c). 
	
	In addition, the similarity distribution between centers and anchor points is larger than that between centers and GD solutions. This ordering suggests that the centers lie deep within the bulk of the solution manifold, with anchor points positioned more peripherally and GD solutions even farther away. 
	This ``nested overlap'' structure has already been observed in the negative perceptron problem, see ref.~\cite{annesi2024exact}. 
	
	Despite the high similarity between the $\boldsymbol{c}$ vectors ($q \approx 0.90 \pm 0.03$), the latter are not linearly connected at fixed norm (Fig.~\ref{fig:Replica}b, upper panel, green curves), even though the barrier height is significantly lower than the one between $\boldsymbol{c}$ and GD solutions $\s$, as well as $\s^{*}$ points (Fig.~\ref{fig:Replica}b, upper panel, blue and red curves, respectively). 
	
	Finally, we performed double--ratchet simulations between pairs of $\boldsymbol{c}$ vectors and between centers and GD solutions. In the latter case, at $T=T_L$ the weight initialized on the center $\boldsymbol{c}$ stays very close to it along the double ratchet dynamics (and its energy within two standard deviations from the average at the simulation temperature); the weight starting from the GD solution instead slowly lowers its energy until the coupled $\boldsymbol{c}$ vector is reached (Fig.~\ref{fig:Replica}c, blue curves).
	
	On the contrary, when the same simulation is performed between pairs of center points at $T<T_L$, the latter are able to connect to each other through paths whose energy is lower than the average $\langle U\rangle_{T_L}$ (Fig.~\ref{fig:Replica}c, green curves).

	From this last computational analysis, we conclude that the center of the spiky low--energy manifold is not barrier--free. Although there are low--energy paths connecting all low--energy regions of the structure, these are not as simple as linear paths. Moreover, there is a small slope of the energy towards the center.

	\subsubsection{The intermediate temperature regime}
	
	By increasing the temperature to intermediate values (cf. Fig. \ref{fig:FixedTempMonteCarlo}), the system reaches a phase where the HMC algorithm can easily sample the space of weights. In this phase the training error is still small but non--negligible (up to $\epsilon\approx 0.3$) but the system can still learn some of the data we present.
	
	The similarity distribution between states shows a single peak centered at $q\approx 0.6$, indicating that most of the sample states are equivalent to each other. This is different from what is found at low temperatures, where there is a difference between points at the periphery of the structure and weights in the center of the low--energy manifold. Such a single peak in $\rho(q)$ is compatible with what is known in the language of complex systems as replica-symmetric behavior of the system~\cite{mezard1987spin}. 
	
	The states sampled during simulations in this intermediate regime ($T=1.8\cdot10^{1}$) show an average similarity $\langle q\rangle=0.54\pm 0.03$ with respect to GD solutions $\s_{i}$ and $\langle q\rangle=0.58\pm 0.04$ with respect to both the $s^*$ and $\boldsymbol{c}$ vectors found in the low--temperature phase.
	Consequently, here the system is never close to any specific region of the subspace explored at low temperatures. In particular, the large mean distance between the center of the manifold and the typical states sampled at intermediate temperatures suggests that the former is entropically unfavorable. This fact, along with the unimodal shape of the cosine--similarity distribution, suggests that the visited manifold at intermediate temperatures still retains a (less rugged) spiky shape, but with a ``hollow'' center characterized by high free energy.
	
	\subsection{The overparametrized regime}\label{sec::overparametrized}
	
	\subsubsection{Similarities and differences with the underparametrized regime}
	
	We then compared the properties of the space of weights that we found at the interpolation threshold with that of the overparametrized regime $\alpha=0.2$ (i.e. $P=200$). The presented results are obtained from different simulations in the stationary regime and averaged over several dataset instances.
	
\begin{figure}
		\centering
		\includegraphics[width=\linewidth]{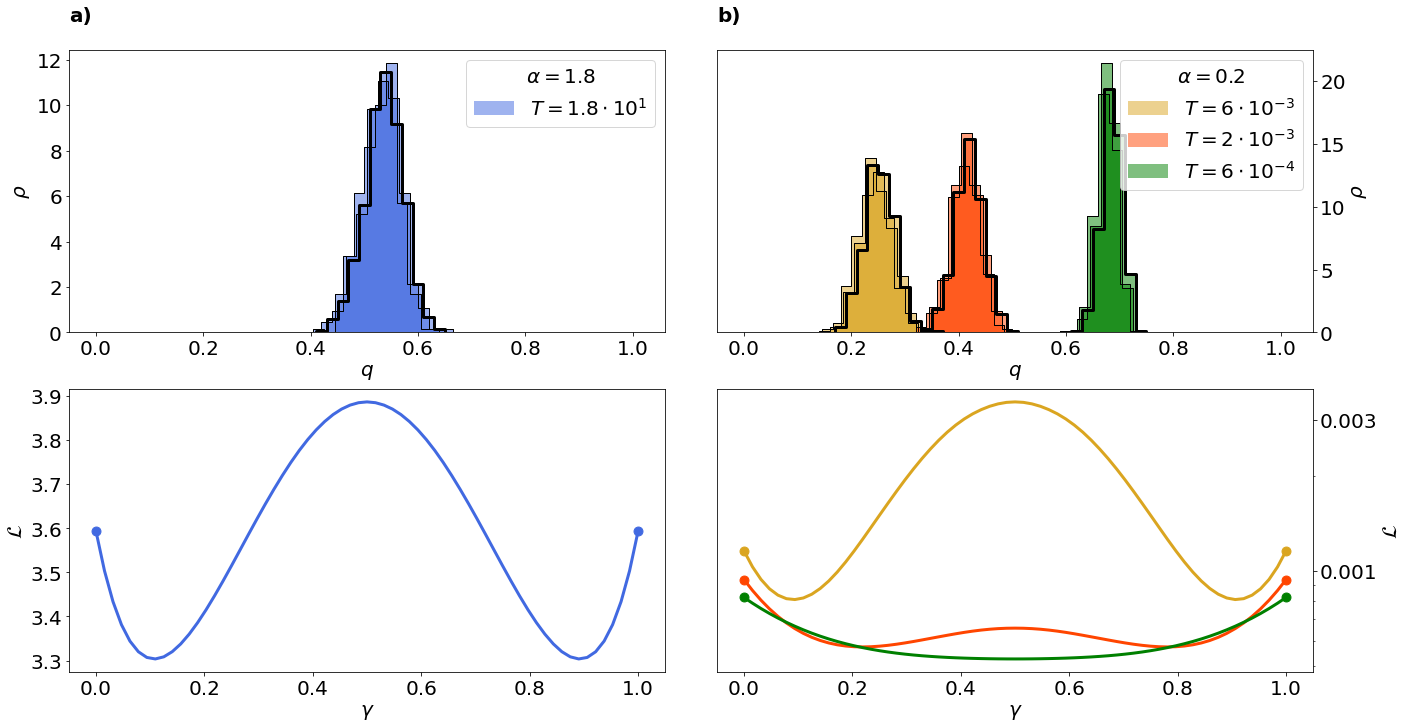}
		\caption{
			\textbf{a)} The thin--line histogram in the upper panel shows two intra-state  overlap distributions $\rho(q)$ associated with two intermediate-temperature ($T=1.8\cdot10^{1}$) simulations starting from different GD solutions close to the interpolation threshold ($\alpha=1.8$). The thick--line histogram stands for the corresponding inter-state distribution computed between the two sets of samples states.
			The three overlapping distributions have been slightly shifted along the $x$--axis to facilitate the comparison. Lower panel: for the same $(\alpha, T)$, we show the binary cross-entropy loss along the mean geodesic curve for points sampled at equilibrium, parametrized by the variable $\gamma \in [0, 1]$.
			\textbf{b)} The same quantities are presented for simulations in the overparametrized regime ($\alpha=0.2$) at $T=6\cdot10^{-3}, 2\cdot10^{-3}, 6\cdot10^{-4}$ (intermediate to low temperatures).
		}
		\label{fig:overparametrized}
	\end{figure}

	We performed various HMC simulations for a wide range of temperatures, all of which never indicate a frozen behavior, as, similarly to the intermediate--temperature case at $\alpha=1.8$, the intra e inter overlap distribution coincide, see Fig. \ref{fig:overparametrized}a. Moreover, the energy profile along the linear paths between the sampled states shows a central free energy barrier, both at $\alpha=1.8$ and at most temperatures at $\alpha=0.2$. In the last case, the central barrier disappears for vanishing temperatures and the energy profile assumes a convex shape (Fig. \ref{fig:overparametrized}b).
	
	In conclusion, the intermediate--temperature configurations visited by the system are similarly distributed both near the interpolation threshold and in the overparametrized regime, where in both scenarios the manifold exhibits a symmetric shape, in the sense that the system populates a single kind of state, belonging to the spiky periphery of the space. Only at low temperatures do the two cases differ, since in the former the explored subspace maintains a spiky shape, where part of the center of the manifold and its periphery are populated, whilst in the latter it becomes convex.
	
	\subsubsection{The barrier along linear paths between sampled solutions can be computed analytically}

	In the overparametrized regime, we can compute analytically both the energy (loss) and the training error along the geodesic path connecting two weights $\boldsymbol{w}^1$, $\boldsymbol{w}^2$ in the thermodynamic limit,
	extending the previous analysis performed in ref.~\cite{Annesi2023star} to the tree committee machine case. 
	In full generality we will assume to sample the two weights from two different Boltzmann distributions $\boldsymbol{w}^1 \sim p_1(\boldsymbol{w}; \mathcal{D})$, $\boldsymbol{w}^2 \sim p_2(\boldsymbol{w}; \mathcal{D})$ whose individual form are the same as in equation~\eqref{eq::Boltzmann} but each of them differing from the choice of the loss function and the prior, i.e.
	\begin{subequations}
		\label{eq::p1_p2}
		\begin{align}
			p_1(\boldsymbol{w}; \mathcal{D}) = \frac{e^{-\beta \mathcal{L}_1(\boldsymbol{w}; \mathcal{D})} p_1(\boldsymbol{w})}{Z^1_{\mathcal{D}}} \\
			p_2(\boldsymbol{w}; \mathcal{D}) = \frac{e^{-\beta \mathcal{L}_2(\boldsymbol{w}; \mathcal{D})} p_2(\boldsymbol{w})}{Z^2_{\mathcal{D}}}
		\end{align}
	\end{subequations}
	Notice, moreover, that in equations~\eqref{eq::p1_p2} the training data $\mathcal{D}$ is the same for both Boltzmann distributions. 
	As in Section~\ref{sec::sampling&connecting} we choose a Gaussian distribution as a prior $p_1(\boldsymbol{w})$, $p_2(\boldsymbol{w})$, with a $L2$ regularization parameter that we denote respectively by $\lambda_1$ and $\lambda_2$. In the large $N$ limit the choice of the regularization parameter will induce a non-trivial value of the norm of the weights $\boldsymbol{w}^1$ and $\boldsymbol{w}^2$. In the following we will suppose that $\lambda_1$ and $\lambda_2$ are chosen such that the norm of $\boldsymbol{w}_1$ is the same as the one of $\boldsymbol{w}^2$, i.e. $|\boldsymbol{w}^1|^2 = |\boldsymbol{w}^2|^2 \equiv Q$. 
	
	We are interested in computing the average training error (remind Eq.~\eqref{eq::training_error} and the discussion in section \ref{sec::model}) and training loss landscape on the geodesic path joining $\boldsymbol{w}^1$ and $\boldsymbol{w}^2$ at fixed squared norm $Q$. This can be obtained as follows. Given $\boldsymbol{w}^1$ and $\boldsymbol{w}^2$ we define the interpolating weight $\boldsymbol{w}(\gamma)$ with parameter $\gamma \in [0,1]$ as the vector whose components are the linear interpolation of the components of $\boldsymbol{w}^1$ and $\boldsymbol{w}^2$
	\begin{equation}
		w_{li}(\gamma) \equiv \gamma w_{li}^1 + (1-\gamma) w_{li}^2\,.
	\end{equation} 
	In order to compute the geodesic path at the same squared norm $Q$, we finally need to project the whole straight path on the hypersphere of radius $\sqrt{Q}$. 
	This defines the weight
	\begin{equation}
		\label{eq::wtildegamma}
		\widetilde{w}_{li}(\gamma) \equiv \frac{w_{li}(\gamma)}{c_\gamma} 
\end{equation}
where 
\begin{multline}
	\label{eq::cgamma}
	c_\gamma \equiv \frac{\lvert \boldsymbol{w}(\gamma) \rvert}{\sqrt{Q}} \equiv \sqrt{\frac{1}{QN} \sum_{l=1}^K \sum_{i=1}^{N/K} w^2_{li}(\gamma)} \\
	= \sqrt{1 - 2\gamma(1-\gamma) \left(1-\frac{p}{Q}\right)}
\end{multline}
In the previous expression we have introduced the quantity
\begin{equation}
	\label{eq::p_def}
	p \equiv \frac{1}{N} \sum_{li} w_{li}^1 w_{li}^2
\end{equation}
which corresponds to the overlap between $\boldsymbol{w}^1$ and $\boldsymbol{w}^2$.

We are interested in finding what is the average value of a loss $\widetilde{\mathcal{L}}$ of the projected interpolated weight $\widetilde{\boldsymbol{w}}(\gamma)$ as a function of $\gamma$, and how this profile depends on the choice of the endpoints $\boldsymbol{w}^1$ and $\boldsymbol{w}^2$ via the probability density functions in~\eqref{eq::p1_p2}. In formulas, what we compute is 
\begin{widetext}
	\begin{equation}
		\label{eq::E(gamma)}
		\begin{split}
			E(\gamma) &= \frac{1}{P} \, \mathbb{E}_{\mathcal{D}} \, 
			\left\langle \widetilde{\mathcal{L}}\left(\widetilde{\boldsymbol{w}}(\gamma); \mathcal{D} \right) 
			\right\rangle_{\boldsymbol{w}^1 \sim p_1(\cdot; \mathcal{D}), \boldsymbol{w}^2 \sim p_2(\cdot; \mathcal{D})} \\
			&= \frac{1}{P} \, \mathbb{E}_{\mathcal{D}} \frac{\int d \boldsymbol{w}^1 d \boldsymbol{w}^2 \, p(\boldsymbol{w}^1) p(\boldsymbol{w}^2) \, e^{-\beta \mathcal{L}_1(\boldsymbol{w}_1; \mathcal{D}) -\beta \mathcal{L}_2(\boldsymbol{w}_2; \mathcal{D})} \, \widetilde{\mathcal{L}}\left(\widetilde{\boldsymbol{w}}(\gamma); \mathcal{D} \right)}{Z_{\mathcal{D}}^1 Z_{\mathcal{D}}^2}
		\end{split}
	\end{equation}
\end{widetext}
In the previous equation we have denoted by $\langle \cdot \rangle$ the average over the Boltzmann distribution~\eqref{eq::Boltzmann} and by $\mathbb{E}_{\mathcal{D}}$ the average over the dataset. Using a general loss $\widetilde{\mathcal{L}}$ allows us also to both have access to training loss and training error profiles. 

Equation~\eqref{eq::E(gamma)} can be computed by the replica method in the large $N$ limit. In the following we will focus only on the infinite width limit $K \to \infty$ with the ratio $K/N \to 0$, as done in~\cite{relu_locent}. The full calculation is reported in Appendix~\ref{sec::theory_calculations_paths}. Here we only mention that the result of the calculation, assuming no replica symmetry breaking of the solution space (i.e. in the so called Replica Symmetric ansatz), will depend on simple geometrical quantities. Those are the typical overlap between two weights $\boldsymbol{w}^a$, $\boldsymbol{w}^b$ extracted from $p_1$, $\boldsymbol{w}^a, \boldsymbol{w}^b \sim p_1(\boldsymbol{w}; \mathcal{D})$ (respectively from $p_2$,  $\boldsymbol{w}^a, \boldsymbol{w}^b \sim p_2(\boldsymbol{w}; \mathcal{D})$), i.e.
\begin{subequations}
	\label{eq::q1_q2}
	\begin{align}
		q_1 &= \mathbb{E}_{\mathcal{D}} \left\langle \frac{1}{N} \sum_{l i} w_{li}^a w_{li}^b \right\rangle_{\boldsymbol{w}^a, \boldsymbol{w}^b \sim p_1(\cdot; \mathcal{D})} \\
		q_2 &= \mathbb{E}_{\mathcal{D}} \left\langle \frac{1}{N} \sum_{li} w_{li}^a w_{li}^b \right\rangle_{\boldsymbol{w}^a, \boldsymbol{w}^b \sim p_2(\cdot; \mathcal{D})} 
	\end{align}
\end{subequations} 
as well as the typical overlap $p$ between the endpoints $\boldsymbol{w}^1$ and $\boldsymbol{w}^2$, which was introduced in equation~\eqref{eq::p_def}. Due to the concentration of the Gibbs measure in the large $N$ limit $p$ can be written as
\begin{equation}
	\label{eq::p}
	p \equiv \mathbb{E}_{\mathcal{D}} \left\langle \frac{1}{N} \sum_{li} w_{li}^1 w_{li}^2 \right\rangle_{\boldsymbol{w}^1 \sim p_1(\cdot; \mathcal{D}), \boldsymbol{w}^2 \sim p_2(\cdot; \mathcal{D})}
\end{equation}
Since the weights $\boldsymbol{w}^1$ and $\boldsymbol{w}^2$ are in general samples of two different probability distributions~\eqref{eq::p1_p2}, the overlaps $q_1$ and $q_2$ can be simply obtained by computing with the replica method the corresponding partition function $Z^1_{\mathcal{D}}$ and $Z^2_{\mathcal{D}}$. We refer to~\cite{relu_locent,Baldassi2023Typical} for such calculation, but we report for convenience of the reader in Appendix~\ref{sec::typical_overlap} its outcome. The overlap $p$ is instead slightly more difficult to compute, because it amounts to study an elastically coupled system of weights $\boldsymbol{w}^1$, $\boldsymbol{w}^2$ in the limit where the coupling is sent to zero, see~\cite{Annesi2023star} for details. We present in Appendix~\ref{sec::overlap_p} an alternative and complementary calculation based on the Franz-Parisi potential~\cite{franz1995recipes}.

\subsubsection{Comparison between theory and simulations}

We compare here our analytical predictions with the numerical results obtained using the HMC algorithm. 
We have considered the case $K/N = 0.005$, a constraint density $\alpha = 0.2$,  temperature $T=0.02$ and L2 regularization $\beta \lambda = 0.02$. 

\begin{figure}[t]
	\includegraphics[width=1\linewidth]{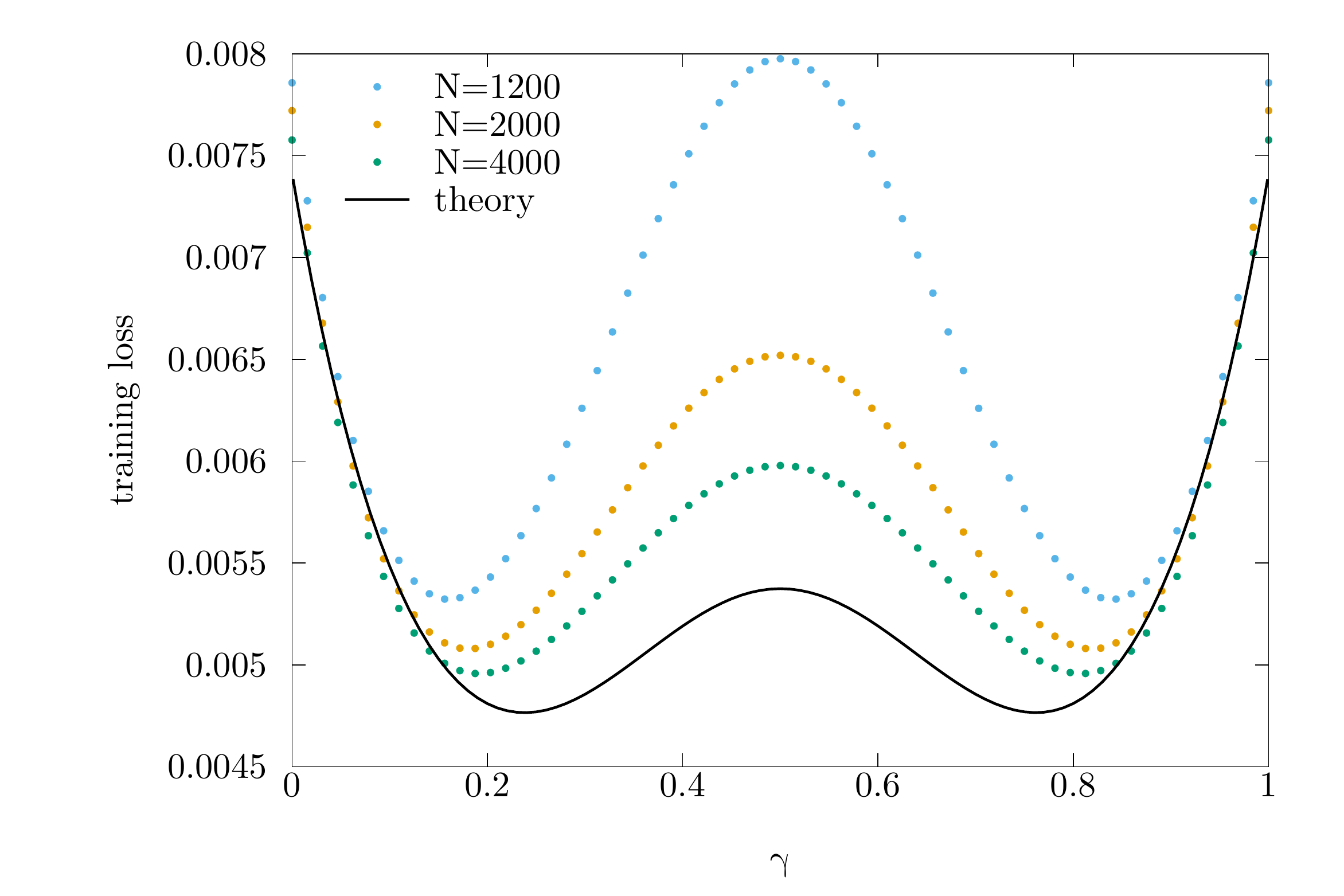}
	\caption{Loss landscape along the (fixed norm) geodesic path interpolating two weight configurations sampled from the Boltzmann distribution~\eqref{eq::Boltzmann} with the cross entropy loss,  temperature $T=0.02$ and regularization parameter $\beta \lambda = 0.02$. The ReLU function is used as the activation function. We have fixed the constrained density to $\alpha = 0.2$ and $K/N = 0.005$, while increasing $N$ (points). Full line refers to the theoretical prediction, which is reported in Eq.~\eqref{eq::anal_result_main} of the appendix.
	}
	\label{fig::comparison}
\end{figure}

The plot of Figure~\ref{fig::comparison} shows the cross-entropy loss along the geodesic path interpolating between two samples of the Boltzmann distribution. We note that the cross-entropy profile presents a non-trivial non-monotonic behavior: starting from one of the two configurations, the loss starts to decrease and then increasing again, reaching a local maximum in the middle of the path. This same non-trivial behavior is also observed in the numerical estimate. We emphasize that achieving the asymptotic limit predicted by the theory is nontrivial, as it requires accounting for finite $N$ and $K$ corrections while operating in the scaling regime $K/N \to 0$. 
Nevertheless, by keeping the ratio $K/N$ small and increasing $N$, we show that the simulations approach the predictions given by our infinite-size theory. 



\subsubsection{The star-shaped property of the solution space}

Thus far, in both the underparametrized and overparametrized regimes we have focused on the energy landscape induced by the cross-entropy loss. Here, we broaden our scope to examine the entire solution space, i.e. we consider the loss function~\eqref{eq::loss_margin}. Notice that this is a larger set of set of weights that includes, but is not limited to, those configurations selected by optimizing the cross-entropy loss. 

\begin{figure}[t]
	\includegraphics[width=0.49\linewidth]{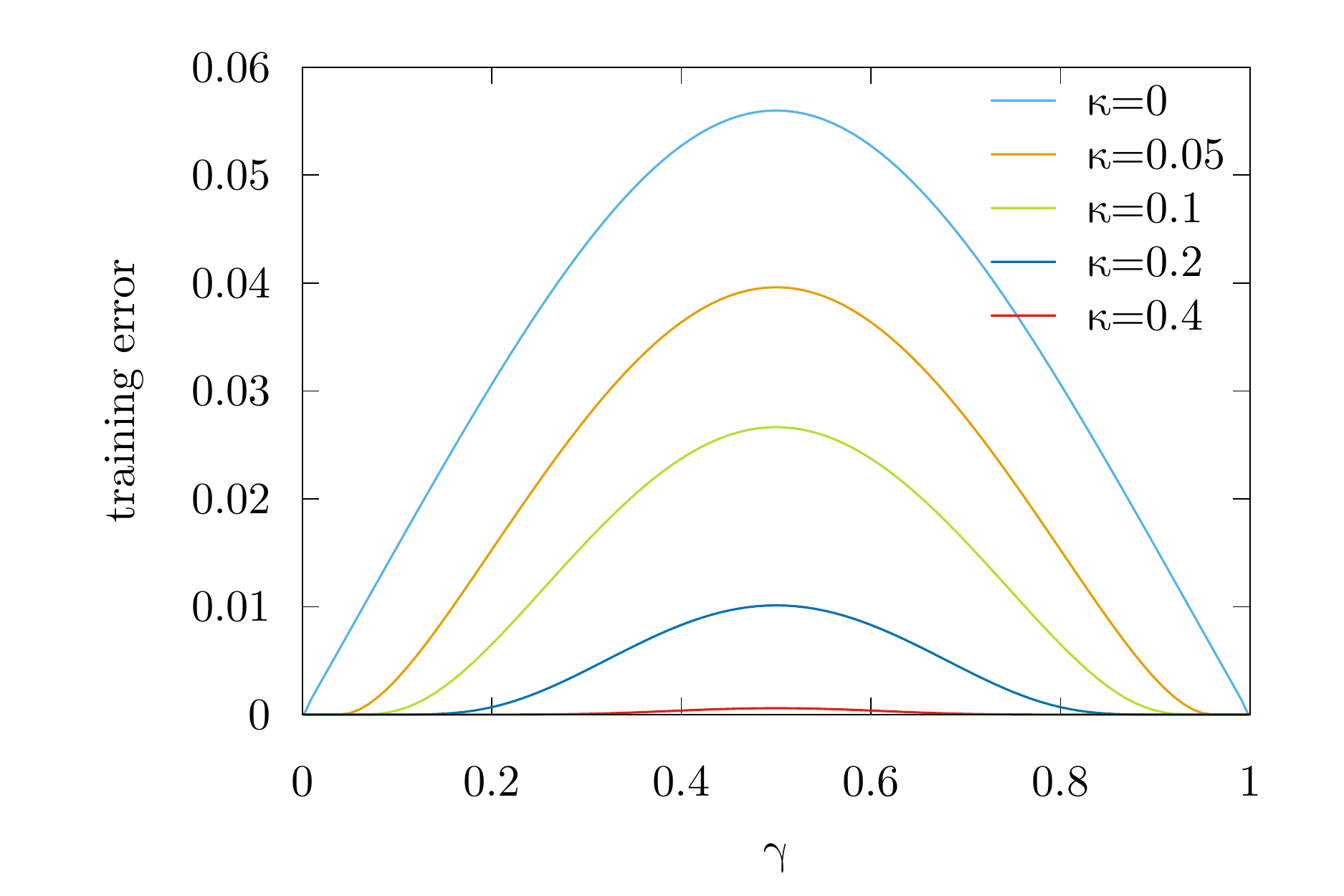}
	\includegraphics[width=0.49\linewidth]{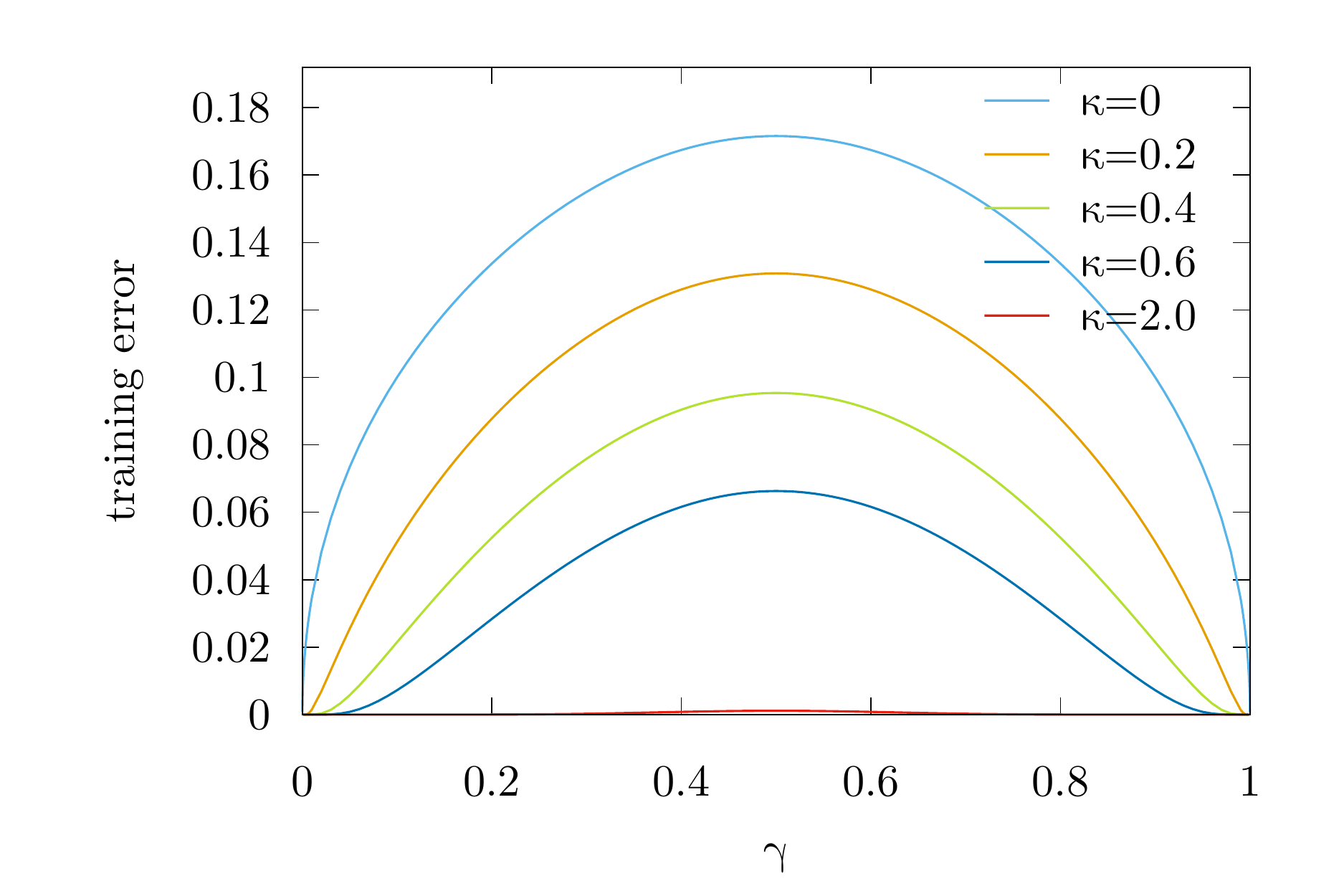}
	\caption{Training error along the geodesic path connecting 2 solutions sampled from the loss function $\ell_r(x) \equiv \Theta(\kappa_r - x)$, $r = 1, 2$ with equal margin $\kappa_1 = \kappa_2 \equiv \kappa$, $\alpha = 0.1$ and with fixed norm $Q=1$. 
		The two panels refers to the choice of the activation function: ReLU activation (left) and sign activation (right panel). The barrier between typical solutions i.e. $\kappa = 0$ is strictly non-vanishing; increasing the margin on the sampled solutions the barrier decreases and eventually vanishes for large enough $\kappa$, similarly to the finding of ref. ~\protect\cite{Annesi2023star}. 
	}
	\label{fig::training_error_typical}
\end{figure}

In Figure~\ref{fig::training_error_typical} we consider the case in which the endpoints are sampled from the large $\beta$ limit of the Boltzmann distributions corresponding to the theta loss~\eqref{eq::loss_margin}, with two equal margin $\kappa_1 = \kappa_2 = \kappa$. We plot the training error ~\eqref{eq::training_error} (i.e. $\widetilde{\mathcal{L}}(\widetilde{\boldsymbol{w}}; \mathcal{D}) = \sum_{\mu = 1}^P \Theta(-y^\mu \Delta^\mu(\boldsymbol{w}))$) on the geodesic path joining such solutions for several values of $\kappa$ in the low $\alpha$ regime, i.e. corresponding to a rather overparameterized network. It can be noticed that  for $\kappa = 0$, i.e. for typical solutions of the learning task, as soon as one moves away from the endpoints the training error is strictly positive, meaning that $\left.\frac{dE(\gamma)}{d\gamma}\right|_{\gamma=0} >0$ and $\left.\frac{dE(\gamma)}{d\gamma}\right|_{\gamma=1} <0$. Increasing the value of $\kappa$ there is a small neighborhood of the endpoints where the training error vanishes. Overall the whole curve of the training error monotonically decreases if one keeps increasing $\kappa$. For $\kappa > \kappa^\star$ the two endpoints become linear mode connected. This has been called in~\cite{Annesi2023star} the ``geodesically convex'' component of the manifold of solutions, as any two solutions sampled within this region are linear mode connected. In the two panels of Figure~\ref{fig::training_error_typical} we also compare two activation functions, the ReLU and the sign activation function. It can be noticed that overall the barrier for a fixed $\kappa$ is smaller in the ReLU activation case. This is consistent with was found in~\cite{relu_locent}, where it has been argued that the training error landscape corresponding to the ReLU activation possess wider and flatter minima with respect to other activation choices like the sign case.

\begin{figure}[t]
	\includegraphics[width=0.49\linewidth]{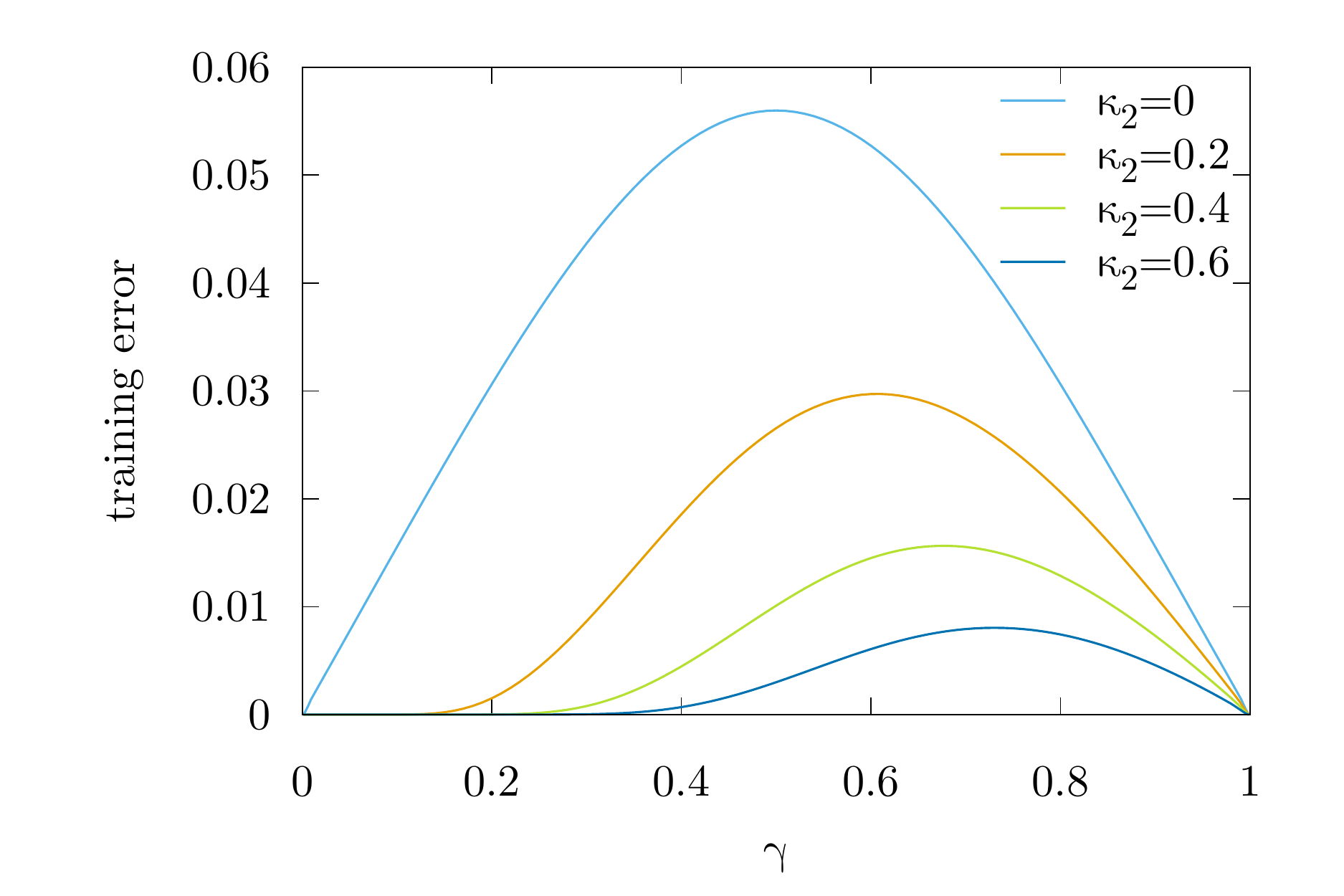}
	\includegraphics[width=0.49\linewidth]{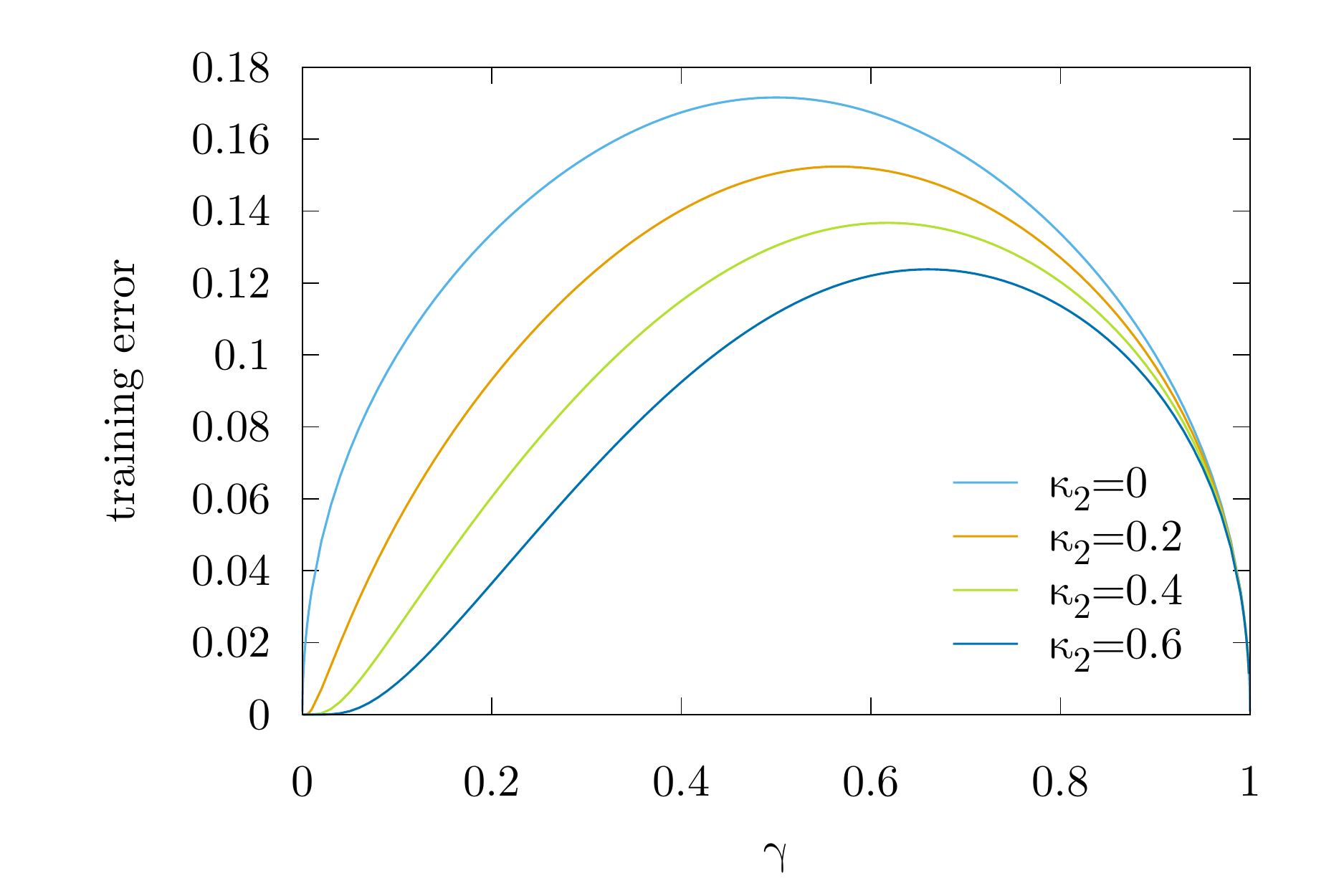}
	\caption{Case where $\ell_r(x) \equiv \Theta(\kappa_r - x)$, $r = 1, 2$ and $\widetilde{\ell}(x) = \Theta(- x)$, with $\alpha = 0.1$ and fixed norm $Q=1$. We here fixed the first endpoint (that is located at $\gamma = 1$) to have a margin $\kappa_1 = 0$. The second endpoint ($\gamma = 0$) has a variable value of the margin $\kappa_2$ (from smaller to larger margin, curves from top to bottom). The left panel corresponds to the ReLU activation, the right one to the sign function. Increasing the robustness $\kappa_2$ the training error barrier on the geodesic monotonically decreases. For large enough $\kappa_2$ the solutions become geodesically connected, as observed in ref. ~\protect\cite{Annesi2023star} for the negative perceptron model. The solution space is therefore star-shaped, namely there exists a set o solutions (those ones with a very large margin), that are geodesically connected to all other solutions with lower margin.
	}
	\label{fig::training_error_atypical}
\end{figure}

In Figure~\ref{fig::training_error_atypical} we consider the case in which the endpoints are sampled from the large $\beta$ limit of the Boltzmann distributions corresponding to the loss~\eqref{eq::loss_margin}, but with two different margin $\kappa_1$ and $\kappa_2$. We consider the first endpoint to be a typical solution of the classification task, i.e. to have margin $\kappa_1 = 0$. Similarly as before, we plot the training error, i.e. $\widetilde{\mathcal{L}}(\widetilde{\boldsymbol{w}}; \mathcal{D}) = \sum_{\mu = 1}^P \Theta(-y^\mu \Delta^\mu(\boldsymbol{w}))$, on the geodesic path joining $\boldsymbol{w}^1$ to $\boldsymbol{w}_2$ for several values of $\kappa_2$ and for the same $\alpha$ considered in Figure~\ref{fig::training_error_typical}. It can be noticed that the maximum of the barrier is always closer to the less robust solution, and the whole curve monotonically decreases if one keeps increasing $\kappa_2$. For $\kappa_2 > \kappa_{\text{krn}}$ the two solutions become eventually linear mode connected. This means that typical solutions despite not being linear mode connected, are connected by a piecewise path, passing through a solution having a rather large margin $\kappa$. It therefore exists a subset of solutions called \emph{kernel}, that are geodesically connected to any other solution of the learning task\footnote{Indeed if $\kappa_2>\kappa_{\text{krn}}$, then not only this solution is linear mode connected to a typical solution with $\kappa_1=0$, but also to any other solution with margin $\kappa_1>0$ extracted from the Boltzmann measure induced by~\eqref{eq::loss_margin}.}. This implies that the space of solutions is \emph{star-shaped} in the overparameterized regime. Similar plots hold for other activation functions; we report the case of the Erf activation function in the appendix. This conclusion is consistent with what was found in reference~\cite{Annesi2023star} for a non-convex but simpler linear model called the negative perceptron. Recently in~\cite{lin2024exploring,sonthalia2024deep}, numerical evidence has been presented suggesting that the solutions space of deep networks possess a star-shaped geometry. We first provide theoretical support for this claim in the case of simple, overparameterized one-hidden-layer networks with general activation functions. Analyzing how this structure evolves in the underparameterized regime would necessitate accounting for Replica Symmetry Breaking effects. 
This study is therefore left to future work.

We refer to Appendix~\ref{sec::xent_errorcounting} for a discussion of the training error and loss along the path connecting a typical solution of the cross entropy loss and the error counting loss, which shows that the low temperature configurations sampled from the cross-entropy loss are solutions located deep into the geodesically convex component of the manifold of solutions. Note that this is consistent with what has been observed numerically in Fig.~\ref{fig:overparametrized}. As we have numerically observed, however, this is not true in the underparametrized regime, as typical cross-entropy loss solutions become linear mode disconnected.

\subsection{The geometrical structure can be robust with respect to highly correlated data} \label{sec::correlated}

\begin{figure*}
	\centerline{\includegraphics[width=\linewidth]{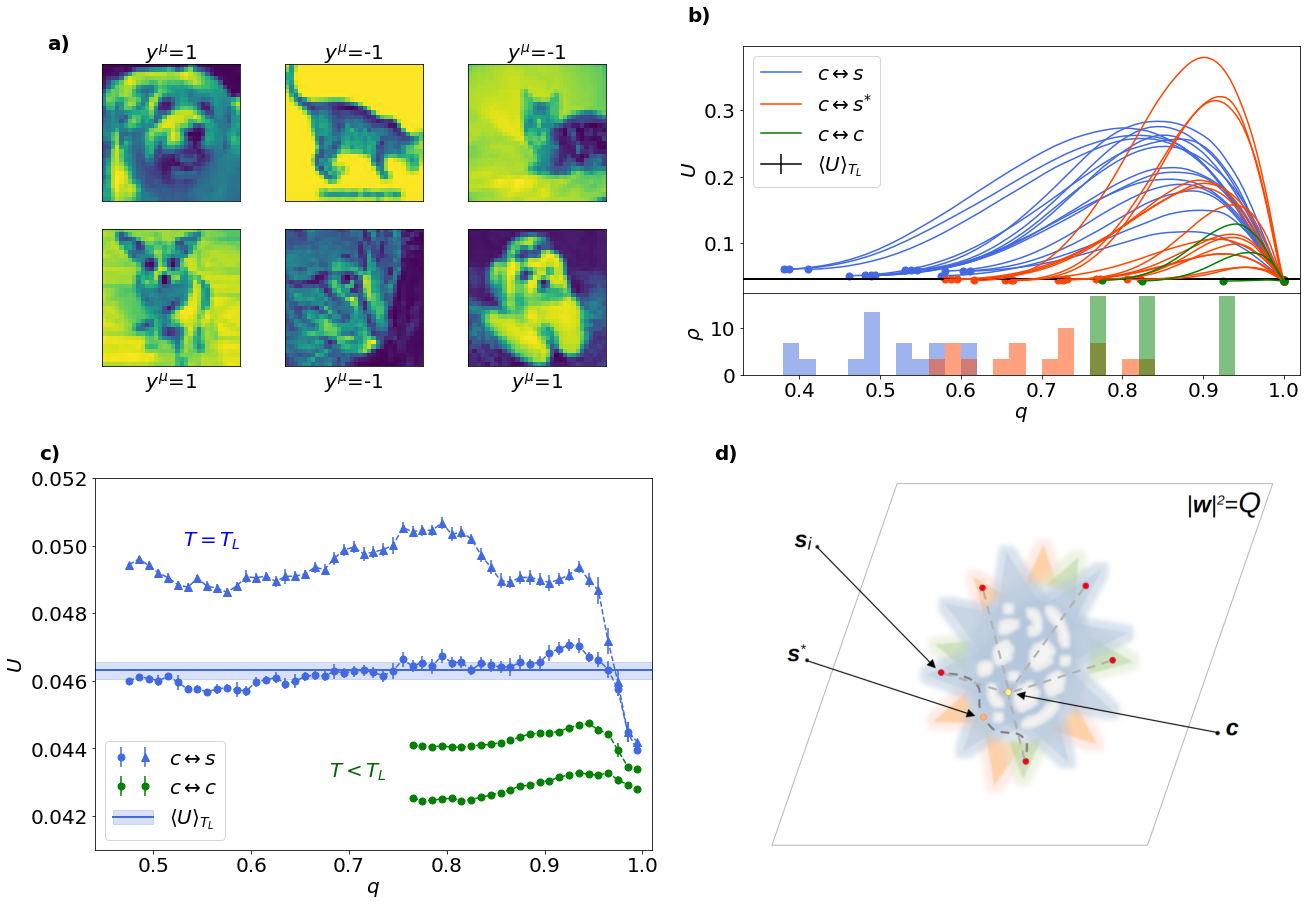}}
	\caption{\textbf{a)} Samples of the images used in the correlated dataset. 
		\textbf{b)} The energy profile along the lines connecting the centers and the GD solutions, between centers and the final points of double ratchet simulations and between centers (upper panel). The black solid line marks the mean energy (and standard deviation) at the simulation temperature $T_L=8.2\cdot 10^{-3}$. The corresponding similarity distribution $\rho(q)$ between centers and solutions (blue bars, lower panel), between centers and double ratchet solutions (orange) and among centers (green). 
		\textbf{c)} The potential energy $U$ as a function of the similarity $q$ for a double--ratchet simulation connecting the center to the GD solution at $T=8.2\cdot 10^{-3}=T_L$ (blue curve) and center to center at $T=2.5\cdot 10^{-3}<T_L$ (green curve). The blue line and shaded area represent respectively the mean energy and standard deviation. 
		\textbf{d)} A sketch of the manifold sampled by the system in the case of correlated data.}
	\label{fig:CorrData}
\end{figure*}

To further assess the applicability of our techniques, we repeated the numerical study on the same architecture, utilizing correlated data for both the training phase and the exploration of the weight space via ratchet and coupled replica simulations. Again, we treated a case of binary classification, where the dataset is composed of 32x32 images of cats and dogs from the CIFAR10 repository, each of which labeled respectively as $+1$ and $-1$. Furthermore, each component of an input vector $\x^{\mu}$, representing a given image in the training dataset, has been scaled so that $x_{i}^{\mu} \in [-1, 1]$ $\forall i, \mu$ (cf. Fig. \ref{fig:CorrData}a).

Due to the nature of the task, we again adopted the binary cross--entropy function with a $L^{2}$ regularization term as the potential energy of the system (see Eq.~(\ref{eq:potentialfunc_impl})), with the Lagrange multiplier $\lambda = 2P \cdot 10^{-8}$. On the contrary, we slightly modified the networks parameters, because of the different size of the input vectors $\x^{\mu}$. In this case, $N/K=K=32$, so that the $i$-th neuron in the hidden layer takes as input the $i$-th row of each input image. Once again, the value of $\alpha = P/N$ is chosen to be just below the threshold where full--batch GD can no longer find weights with zero training error. In this case, that is $\alpha = 0.8$ (or equivalently $P\approx820$).

Similarly to the case of random inputs, the energy of the sampled state decreases as we move from the GD solutions to the center (upper panel in Fig. \ref{fig:CorrData}b). 
An apparent difference between the two cases is the increase in the overlap between the similarity distributions computed among $c$ states and among the anchor points $\s^\star$ found by the double ratchet to connect the GD solutions (lower panel in Fig. Fig.~\ref{fig:CorrData}b). Also, the distribution of similarities between center vectors spreads over lower values as compared to the random--input case (lower panel in Fig. Fig.~\ref{fig:CorrData}b and lower panel in Fig. Fig.~\ref{fig:Replica}b, respectively). This suggests that the center of the spiky manifold is wider and closer to the out-of-equilibrium GD solutions for correlated data.

Moreover, the energy profile seems to display higher peaks and to be more rugged than the random label case, as apparent from the shape of the energy along the paths generated by double ratchet simulations connecting $\boldsymbol{s}$ and $\boldsymbol{c}$ vectors (Fig. \ref{fig:CorrData}c, blue lines) and from the irregular relation between the height of the peaks found along the linear paths and the position in the solutions manifold (compare the upper panel of Fig. \ref{fig:CorrData}b with that of Fig. \ref{fig:Replica}b).

The central points are still connected by low--energy paths (Fig.~\ref{fig:CorrData}c), even though the mean similarity between them is lower than in the random--label case ($\langle q \rangle \approx 0.84 \pm 0.06$).

Summing up, the low--energy states of a tree committee machine trained with correlated data still have a spiky shape, but its energy profile is more rugged and irregular than the random--label case. Moreover, the center is more bulky (see sketch in Fig. \ref{fig:CorrData}d).

\section{Discussion and Conclusions}~\label{sec::conclusions}

In this work, we analyzed the loss landscape of a simple one-hidden-layer artificial neural network with a tree-like structure in both the underparameterized and overparameterized regimes. We employed various numerical techniques, including (Hamiltonian) Monte Carlo methods and biased dynamics that were originally developed in statistical mechanics and have been used to identify native-like conformations in protein folding molecular dynamics simulations. This approach allowed us to sample the weight space manifold at different loss levels and investigate low-energy paths connecting distinct weights.



Close to the interpolation threshold, the numerical exploration of the weight space by using HMC starting from GD solutions, identified two main regimes as a function of temperature. At low temperatures, the system is in a frozen state having zero training error and is unable to move sufficiently far away from initialization (the inter-state overlap distribution is peaked near 1), despite the overlap between HMC trajectories starting from different GD solutions (intra-state overlap) is strictly less then 1. The linear interpolation between GD solutions also shows a loss barrier. Despite this, GD solutions can be connected by a low energy path; those paths are tortuous and difficult to find by an unbiased Monte Carlo algorithm. We identified them by using a double--ratchet hybrid Monte Carlo algorithm, which penalizes moves that cause the two gradient descent solutions to drift apart. These results suggest that the manifold of low-energy weights has a spiky topology, 
with gradient descent solutions located along its protruding rays. 
We have also shown that the center of the manifold solutions has also has a complex pattern of valleys and barriers. 


At intermediate temperatures, the training error is small but not zero. Differently from the low temperature regime, the HMC dynamics is ergodic, since there is no difference between inter- and intra-state overlaps. This means that the shape of the populated state is particularly symmetric and corresponds, in the language of replica calculations, to a replica--symmetric solution. The energy always displays a maximum at the center of the straight lines between states populated at intermediate temperatures. 

The symmetric, hollow structure of states at intermediate temperature is similar to that displayed by the system in the overparametrized regime, suggesting that the two manifolds are quite similar. The main difference between the overparametrized regime and that at the interpolation threshold is at low temperature; while in the latter case we have the spiky shape of the manifold discussed above, in the former the solutions are more spherically distributed, and located at the center of the manifold.

In the overparametrized regime we have also resorted to replica computation to study both the loss and the training error landscape on the linear interpolation between two weights sampled with different Boltzmann probability distributions. Our work shows that the training error manifold is star-shaped: it exists a subset of robust, solutions having a large margin $\kappa$ that are linear mode connected to solutions having lower (even zero) margin, similarly to what was found in ref. ~\cite{Annesi2023star}. Typical low temperature weights extracted from the Boltzmann measure equipped with the cross entropy loss function tend to focus on the inner core of the star-shaped manifold, that we called geodesically convex component, as any two solutions in this region are linear mode connected. This result also agrees with numerical simulations. Our theory also reproduces the non-monotonic behavior of the energy along the linear interpolation between Boltzmann samples at larger temperatures. 


The use of a realistic training dataset, displaying correlated data, instead of random data, does not change substantially the properties of the space of weights at low temperature. The center of the spiky structure becomes more bulky and the barriers more heterogeneous, but the overall geometry does not change.

\vspace{0.5cm}
\noindent{\bf Code availability}

The code used for the numerical calculations is freely available on \href{https://github.com/guidotiana/nnSampling}{Github}.


\bibliography{references.bib}

\clearpage


\onecolumngrid 



\appendix

\section{Equilibrium measure} \label{sec::typical_overlap}

We report here the outcome of the equilibrium calculation, through the evaluation of the free entropy
\begin{equation}
	\phi = \lim_{N\to \infty} \frac{1}{N} \mathbb{E}_{\mathcal{D}} \ln Z(\beta; \mathcal{D})
\end{equation}
where is the partition function defined in equation~\eqref{eq::partition_function} with the Gaussian prior~\eqref{eq::prior}; $\mathbb{E}_{\mathcal{D}}$ refers to the average over the random dataset $\mathcal{D}$. This average can be performed by using the replica trick~\cite{mezard1987spin} and the saddle point method in the large $N$ limit. Here we report the final result, which is a slight variation of the one reported in~\cite{relu_locent}
\begin{equation}
	\label{eq::phi_eq}
	\phi = \lim_{N\to \infty} \lim_{n\to 0} \frac{1}{nN } \ln \mathbb{E}_{\mathcal{D}} Z^n = \mathrm{extr}_{q, Q} \left[ \mathcal{G}_{S}(q, Q) + \alpha \mathcal{G}_{E}(q, Q) \right]
\end{equation}
where we have defined the entropic and the energetic terms respectively as
\begin{equation}
	\begin{split}
		\mathcal{G}_{S}(q, Q) &= \frac{Q}{2(Q-q)} -  \frac{\beta \lambda}{2} Q + \frac{1}{2} \ln(2\pi (Q-q)) \\ 
		\mathcal{G}_E(q, Q) &= \int \prod_{l=1}^K Dx_l \ln \int \prod_{l=1}^K D\lambda_l \, 
		e^{-\beta \ell\left(\frac{1
			}{\sqrt{K}} \sum_l c_l g\left( \sqrt{q} x_l + \sqrt{Q-q} \lambda_l \right) \right)}
	\end{split}
\end{equation}
We remind that $\lambda$ is the Lagrange multiplier (or regularization in machine learning jargon) that fixes the square norm $Q$ of the weights $\boldsymbol{w}$. $q$ and $Q$ can be obtained from the extremization of the right side of~\eqref{eq::phi_eq}. $q$ represent the typical overlap of two weights $\boldsymbol{w}^a$, $\boldsymbol{w}^b$ sampled from the Boltzmann distribution~\eqref{eq::Boltzmann} (see also equation~\eqref{eq::q1_q2}); $Q$ represent the typical squared norm of a sample of~\eqref{eq::Boltzmann}.
In the large $K$ limit (but having $K/N \to 0$), the energetic term can be simplified by using the central limit theorem as shown in~\cite{relu_locent,annesi2024exact}
\begin{equation}
	\begin{split}
		\mathcal{G}_E(q, Q) 
		&=  \int Dz_0 \ln \int Dz_1 \, e^{-\beta \ell\left(\sqrt{\Delta_Q(q) - \Delta_Q(0)} z_0  + \sqrt{\Delta_Q(Q)-\Delta_Q(q)} z_1 - \kappa \right)} \,.
	\end{split}
\end{equation}
where
\begin{equation}
	\Delta_Q(q) \equiv \int Dx \left[ \int Dy \, \varphi\left( \sqrt{q} x + \sqrt{Q-q} y \right) \right]^2 
\end{equation}
is an effective order parameter~\cite{barkai1990statistical,engel1992storage} whose expression depends on the choice of the activation function $\varphi$. This kernel has also the same expression of the Neural Network Gaussian Process (NNGP) kernel that appears in neural networks learning a finite number of examples in the large width limit~\cite{Neal1996}.

\subsection{Large $\beta$ limit}

In the large $\beta$ limit we have the following scaling 
\begin{equation}
	q = Q - \frac{\delta q}{\beta}
\end{equation}
This induces the following scaling on the effective order parameter difference
\begin{equation}
	\Delta_Q(Q) - \Delta_Q(q) \simeq \Delta'_Q(Q) \frac{\delta q}{\beta}
\end{equation}
where
\begin{equation}
	\label{eq::effective_order_parameter_derivative}
	\Delta_Q'(q) \equiv  \frac{\partial \Delta_Q(q)}{\partial q} = \int Dx \left[ \int Dy \, \varphi'\left(\sqrt{q} x + \sqrt{Q-q} y\right) \right]^2\,.
\end{equation}
We therefore have that the free energy of the system is
\begin{subequations}
	\begin{align}
		-f &\equiv \lim_{\beta \to \infty} \frac{\phi}{\beta} = \mathrm{extr}_{\delta q, Q} \left[ \mathcal{G}_{S}(\delta q, Q) + \alpha \mathcal{G}_{E}(\delta q, Q) \right] \\
		\mathcal{G}_{S}(\delta q, Q) &= \frac{Q}{2\delta q} -  \frac{\lambda}{2} Q \\ 
		\mathcal{G}_E(\delta q, Q) &= \int Dz_0 \, z_\star(z_0) \,,
	\end{align}
\end{subequations}
where we have defined the function $z_\star(z_0)$ as
\begin{equation}
	\label{eq::zstar}
	\begin{split}
		z_\star(z_0) &= \argmax_{z_1} \left[ -\frac{z_1^2}{2} - \ell \left( \sqrt{\Delta_Q(Q) - \Delta_Q(0)} \, z_0 + \sqrt{\Delta_Q'(Q) \delta q} \, z_1 \right) \right]
	\end{split}
\end{equation}

\section{Loss Landscape on the linear interpolation between weights} ~\label{sec::theory_calculations_paths}

In this section we want to compute the average loss $\widetilde{\mathcal{L}}$ of $\widetilde{\boldsymbol{w}}(\gamma)$ as defined in~\eqref{eq::E(gamma)}. $\widetilde{\boldsymbol{w}}(\gamma)$ is the interpolation of $\boldsymbol{w}^1$ and $\boldsymbol{w}^2$, see equation~\eqref{eq::wtildegamma}. As stated in the main text, both weights $\boldsymbol{w}^1$ and $\boldsymbol{w}^2$ and their interpolation $\widetilde{\boldsymbol{w}}(\gamma)$ are considered to have the same squared norm $Q$. Equation~\eqref{eq::E(gamma)} can be written also in terms of the corresponding loss per pattern $\widetilde{\ell}$
\begin{equation}
	\begin{split}
		E(\gamma)
		&= \mathbb{E}_{\mathcal{D}} \frac{\int d \boldsymbol{w}^1 d    \boldsymbol{w}^2 \, p(\boldsymbol{w}^1) p(\boldsymbol{w}^2) \, e^{-\beta \mathcal{L}_1(\boldsymbol{w}_1; \mathcal{D}) -\beta \mathcal{L}_2(\boldsymbol{w}_2; \mathcal{D})} \, \tilde \ell\left(\Delta^{\mu_\star}(\widetilde{\boldsymbol{w}}(\gamma)) \right)}{Z_{\mathcal{D}}^1 Z_{\mathcal{D}}^2}
	\end{split}
\end{equation}
we have here focused on a particular pattern $\mu_\star$, since in the thermodynamic limit all input patterns will give the same contribution on average.

\subsection{Replica approach}

The computation proceeds as usual introducing replicas via the identity $(Z_{\mathcal{D}}^r)^{-1} = \lim\limits_{n \to 0} (Z_{\mathcal{D}}^{r})^{n-1}$ with $r = 1, 2$. In the following we avoid explicating the range of each sum and product for simplicity and to lighten up notation. We however specify here that the indexes $r, s = 1, 2$ will run over the real replicas $\boldsymbol{w}^1$ and $\boldsymbol{w}^2$, the index $\mu = 1, \dots, P$ will denote the data index, the indexes $a, b = 1, \dots, n$ will run over the $n$ replicas of the system, the index $l = 1, \dots, K$ will run over the $K$ hidden units and finally $i=1, \dots, N/K$ will be the index running over the weights in single branch of the tree network. 

We have
\begin{equation}
	\label{eq::first_step_replica}
	\begin{split}
		E(\gamma) &= \int \prod_{a r l \mu} \frac{d v_{lr}^{\mu a} d \hat v_{lr}^{\mu a}}{2\pi} e^{i v_{lr}^{\mu a}\hat v_{lr}^{\mu a}} \int \prod_{a r} d \boldsymbol{w}^{r a} \, \prod_{ra\mu} e^{- \beta \ell_r \left(\frac{1}{\sqrt{K}} \sum_l c_l \varphi \left(v^{\mu a}_{lr}\right)\right)}  \, \tilde \ell\left[ \frac{1}{\sqrt{K}} \sum_l c_l \, \varphi\left( \frac{\gamma v_{l 1}^{\mu_\star 1} + (1-\gamma) v_{l 2}^{\mu_\star 1}}{c_\gamma} \right) \right] \\
		&\times e^{- \frac{\lambda_1}{2} \sum_{li a} (w_{li}^{ar=1})^2 - \frac{\lambda_2}{2} \sum_{li a} (w_{li}^{ar=2})^2} \prod_{l i \mu} \mathbb{E}_{\xi_{li}^\mu} e^{-i \frac{\xi_{li}^\mu}{\sqrt{N}} \sum_{a r} w_{li}^{ar} \hat v_{lr}^{\mu a} } \\
		&= \int \prod_{ar, bs, l} \frac{d q_{rsl}^{ab} d \hat q_{rsl}^{ab}}{2\pi} e^{- N \sum_{ar, bs l} q_{rsl}^{ab} \hat q_{rsl}^{ab} - \frac{\lambda_1}{2} \sum_{l a} q^{aa}_{11l} - \frac{\lambda_2}{2} \sum_{rl a} q^{aa}_{22l} + N G_S + (N\alpha - 1) G_E} \\
		&\times
		\int \prod_{a r l} \frac{d v_{lr}^{a} d \hat v_{lr}^{a}}{2\pi} \, e^{i v_{lr}^{a}\hat v_{lr}^{a}} \prod_{ra} e^{- \beta \ell_r \left(\frac{1}{\sqrt{K}} \sum_l c_l \varphi \left(v^{a}_{lr}\right)\right)}  \, \tilde \ell\left[ \frac{1}{\sqrt{K}} \sum_l c_l \, \varphi\left( \frac{\gamma v_{l 1}^{1} + (1-\gamma) v_{l 2}^{1}}{c_\gamma} \right) \right]  e^{-\frac{1}{2} \sum_{ar, bs l}q^{ab}_{rsl} \hat v_{lr}^{a} \hat v_{ls}^{b}}
	\end{split}
\end{equation}
where we remind that $c_\gamma$ is the the quantity that appears in equation~\eqref{eq::wtildegamma}. In the second equality of the previous equation we have performed the average over the Gaussian input data and introduced the following overlaps quantities
\begin{equation}
	q^{ab}_{rsl} \equiv \frac{K}{N} \sum_{i=1}^{N/K} w^{ar}_{li} w^{bs}_{li}
\end{equation}
via delta functions and their integral representations. This enables the decoupling over the site index $i = 1, \dots, N/K$ and the ($P-1$) pattern index $\mu \ne \mu_\star$.  
The last line of~\eqref{eq::first_step_replica} refers to the contribution coming from the pattern index $\mu_\star$ whose dependence we have dropped for simplicity. 
We have also introduced the usual entropic and energetic terms~\cite{relu_locent} as
\begin{equation}
	\begin{split}
		G_S &\equiv \frac{1}{K}\ln\int \prod_{arl} dw^{ra}_l \, e^{\sum_{ar, bs l} \hat q^{ab}_{rs l } w_l^{ra} w_l^{sb}} \\
		G_E &\equiv \ln \int \prod_{a r l } \frac{d v_{rl}^{a} d \hat v_{rl}^{a}}{2\pi} \, e^{-\beta \sum_{r a}\ell_r\left(\frac{1}{\sqrt{K}} \sum_l c_l \, \varphi\left( v^{a}_{rl} \right) \right)} \, e^{i \sum_{rla} v_{rl}^{a}\hat v_{rl}^{a} -\frac{1}{2} \sum_{ar, bs l} q^{ab}_{rs l} \hat v_{rl}^{a} \hat v_{sl}^{b} } \,.
	\end{split}
\end{equation}

\subsection{Replica Symmetric ansatz}
We impose the Replica Symmetric (RS) ansatz over order parameters
\begin{subequations}
	\begin{align}
		q_{rs l}^{aa} &\equiv Q\delta_{rs} + (1-\delta_{rs}) p &  a\in [n]\,,  \forall l\\
		q_{rs l}^{ab} &\equiv t_{rs} = q_r \delta_{rs} + (1 - \delta_{rs}) p & a\ne b\,, \forall l
		\label{eq::trs}
	\end{align}
\end{subequations}
Notice that we called by $q_1$ and $q_2$ the typical overlap between solutions extracted from the distribution of the endpoint $\gamma = 1$ and $\gamma=0$ respectively. The overlap $p$ represents the typical overlap between the two endpoints. A similar ansatz is imposed over the conjugated order parameters $\hat q_{rsl}^{ab}$. 
However in the $n\to 0$ limit the conjugated order parameters will not appear explicitly in the expression of $E(\gamma)$, as the first line after the equal sign in~\eqref{eq::first_step_replica} goes to 1 in the small $n$ limit. We need to express decompose the term
\begin{equation}
	\begin{split}
		-\frac{1}{2} \sum_{abrs l} q_{rs l}^{ab} \hat{v}_{lr}^a \hat{v}_{ls}^b &= -\frac{1}{2} \sum_{r} (Q-q_{r}) \sum_{al} \left(\hat{v}_{rl}^a\right)^2 - \frac{1}{2} \sum_{rs} t_{rs} \sum_{ab l} \hat{v}_{rl}^a \hat{v}_{sl}^b \\
		&= -\frac{1}{2} \sum_{r} (Q-q_{r}) \sum_{al} \left(\hat{v}_{rl}^a\right)^2 - \frac{1}{2} \sum_{rl} \left( \sum_{as} \mathcal{T}_{rs} \hat{v}^a_{sl} \right)^2
	\end{split}
\end{equation}
where $\mathcal{T}_{rs}$ is the $(r,s)$ element of the \emph{square root} of the matrix $t_{rs}$ defined in~\eqref{eq::trs}. The computation proceeds in a standard way by using a Hubbard-Stratonovich transformation and integrating over $\hat{v}_{rl}^a$. We get
\begin{equation}
	\label{eq::last_before_largeK}
	\begin{split}
		E(\gamma)
		&=\int \prod_{a r l} \frac{d v_{rl}^{a} d \hat v_{rl}^{ a}}{2\pi} \, e^{ -\beta \ell_r \left(\frac{1}{\sqrt{K}} \sum_l c_l \varphi \left(v^{a}_{lr}\right) \right)} \, \tilde \ell \left[ \frac{1}{\sqrt{K}} \sum_l c_l \, \varphi\left( \frac{\gamma v_{l 1}^{1} + (1-\gamma) v_{l 2}^{1}}{c_\gamma} \right) \right] \\
		&\times e^{i \sum_{arl}v_{rl}^{a}\hat v_{rl}^{a} -\frac{1}{2} \sum_{r} (Q-q_{r}) \sum_{al} \left(\hat{v}_{rl}^a\right)^2 - \frac{1}{2} \sum_{rl} \left( \sum_{as} \mathcal{T}_{rs} \hat{v}^a_{sl} \right)^2} \\
		&= \int \!\prod_{rl} Dx_{rl} \frac{\int \prod_{rl} \frac{d v_{rl} d \hat v_{rl}}{2\pi} \, e^{i \hat v_{rl} \left( v_{rl} + \sum_s \mathcal{T}_{rs} x_{sl}\right) - \frac{1}{2} (Q-q_r) \hat v_{rl}^2 - \beta \sum_r \ell_r \left(\frac{1}{\sqrt{K}} \sum_l c_l \varphi \left(v_{lr}\right) \right)} \, \tilde\ell\left[ \frac{1}{\sqrt{K}} \sum_l c_l \, \varphi\left( \frac{\gamma v_{l1} + (1-\gamma) v_{l 2}}{c_\gamma} \right) \right]}{\prod_r \int \prod_l \frac{d v_l d \hat v_l}{2\pi} \, e^{ -\beta \ell_r \left(\frac{1}{\sqrt{K}} \sum_l c_l \varphi \left(v_{l}\right) \right)} \, e^{i \sum_l \hat v_l \left( v_l + \sum_s \mathcal{T}_{sr} \, x_{rl } \right) - \frac{1}{2} \sum_r (Q-q_r) \sum_l \hat v_{l}^2 } } \\
		&= \int \!\prod_{rl} Dx_{rl} \frac{\int \prod_{rl} D v_{rl} \, e^{- \beta \sum_r \ell_r \left(\frac{1}{\sqrt{K}} \sum_l c_l \varphi \left(\sqrt{Q-q_r} v_{lr} - \sum_s \mathcal{T}_{rs} x_{sl} \right) \right)} \, \tilde\ell\left[ \frac{1}{\sqrt{K}} \sum_l c_l \, \varphi\left( 
			\sum_r \gamma_r \left( \frac{\sqrt{Q-q_r}v_{rl} - \sum_s \mathcal{T}_{rs} x_{sl}}{c_{\gamma}} \right) \right) \right] }{\prod_r \int \prod_l Dv_l \, e^{ -\beta 
				\ell_r \left(\frac{1}{\sqrt{K}} \sum_l c_l \varphi \left( \sqrt{Q-q_r} v_{l} - \sum_s \mathcal{T}_{rs} x_{sl}  \right) \right)} }
	\end{split}
\end{equation}
where in the last expression we have defined $\gamma_1 \equiv \gamma$ and $\gamma_2 = 1-\gamma$ for convenience. 

\subsection{Large $K$ limit}

We now perform the large $K$ limit, in order to further simplify and the expression in~\eqref{eq::last_before_largeK} by repeated usage of the central limit theorem. We will call by $I$ the numerator of the last expression in~\eqref{eq::last_before_largeK}.

The numerator of the fraction can be written as
\begin{equation}
	\label{eq::largeK_step1_first}
	\begin{split}
		I
		&\equiv \int \prod_{rl} D v_{rl} \, e^{- \beta \sum_r \ell_r \left(\frac{1}{\sqrt{K}} \sum_l c_l \varphi \left(\sqrt{Q-q_r} v_{lr} - \sum_s \mathcal{T}_{rs} x_{sl} \right) \right)} \, \tilde\ell\left[ \frac{1}{\sqrt{K}} \sum_l c_l \, \varphi\left( 
		\sum_r \gamma_r \left( \frac{\sqrt{Q-q_r}v_{rl} - \sum_s \mathcal{T}_{rs} x_{sl}}{c_{\gamma}} \right) \right) \right] \\
		&= \int \prod_r \frac{d h_r d \hat h_r}{2\pi} \, e^{i \sum_r h_r \hat h_r -\beta \sum_r \ell_r\left( h_r \right)} \int \frac{d \eta d \hat \eta}{2\pi} e^{i \hat \eta \eta} \, \tilde{\ell}\left[ \eta \right] \\
		&\times \int \prod_{rl} D v_{rl} \, e^{-\frac{i}{\sqrt{K}}\sum_r \hat h_r \sum_l c_l \varphi \left(\sqrt{Q-q_r} v_{lr} - \sum_s \mathcal{T}_{rs} x_{sl} \right) -\frac{i \hat \eta}{\sqrt{K}} \sum_l c_l \, \varphi\left( \sum_r \gamma_r \left( \frac{\sqrt{Q-q_r}v_{rl} - \sum_s \mathcal{T}_{rs} x_{sl}}{c_{\gamma}} \right) \right)} 
	\end{split}
\end{equation}
Expanding the exponential up to second order, averaging over $v_{lr}$ and re-exponentiating we get
\begin{equation}
	\label{eq::largeK_step1_reexp}
	\begin{split}
		&\int \prod_{rl} D v_{rl} \, e^{-\frac{i}{\sqrt{K}}\sum_r \hat h_r \sum_l c_l \varphi \left(\sqrt{Q-q_r} v_{lr} - \sum_s \mathcal{T}_{rs} x_{sl} \right) -\frac{i \hat \eta}{\sqrt{K}} \sum_l c_l \, \varphi\left( \sum_r \gamma_r \left( \frac{\sqrt{Q-q_r}v_{rl} - \sum_s \mathcal{T}_{rs} x_{sl}}{c_{\gamma}} \right) \right)}  \\
		&\simeq e^{- i \sum_r \hat h_r M_r^{(0)} - i \hat \eta N^{(0)} - \frac{1}{2} \sum_{r} \Delta_r^{(0)} \hat h_r^2 - \frac{\Xi^{(0)}}{2} \hat \eta^2 - \hat \eta \sum_r \hat h_r \Omega_r^{(0)}}
	\end{split}
\end{equation}
where we have introduced the notation
\begin{subequations}
	\begin{align}
		\varphi_{rl}(v_r, x) &\equiv \varphi \left(\sqrt{Q-q_r} v_{lr} - \sum_{s} \mathcal{T}_{rs} x_{sl} \right) \\
		\tilde{\varphi}_l(\gamma; \{ v_r \}_{r=1}^y, x) &\equiv \varphi\left(\sum_r \gamma_r \left( \frac{\sqrt{Q-q_r}v_{rl} - \sum_s \mathcal{T}_{rs} x_{sl}}{c_{\gamma}} \right) \right)
	\end{align}
\end{subequations}
to define the following quantities
\begin{subequations}
	\begin{align}
		M^{(0)}_r &\equiv \frac{1}{\sqrt{K}} \sum_l c_l \, \left\langle \varphi  \left(\sqrt{Q-q_r}v_r  - \sum_{s} \mathcal{T}_{rs} x_{sl}  \right)  \right\rangle_{v_r} \equiv \frac{1}{\sqrt{K}} \sum_l c_l \, \left\langle \varphi_{rl}  \right\rangle_{v} \\
		N^{(0)} &\equiv \frac{1}{\sqrt{K}} \sum_l c_l \, \left\langle \varphi \left(\sum_r \gamma_r \left( \frac{\sqrt{Q-q_r}v_{rl} - \sum_s \mathcal{T}_{rs} x_{sl}}{c_{\gamma}} \right)\right) \right\rangle_{\{v_r \}_{r=1}^y} \equiv \frac{1}{\sqrt{K}} \sum_l c_l \, \left\langle \tilde{\varphi}_l \right\rangle_{v}\\
		\Delta^{(0)}_r &\equiv \frac{1}{K} \sum_l c_l^2 \, \left[ \langle \varphi^2_{rl}\rangle_{v} - \langle \varphi_{rl}\rangle^2_{v} \right] \\
		\Xi^{(0)} &\equiv \frac{1}{K} \sum_l c_l^2 \left[ \langle \tilde \varphi^2_{l}\rangle_{v} - \langle \tilde \varphi_{l}\rangle^2_{v}  \right] \\
		\Omega_r^{(0)} &\equiv \frac{1}{K} \sum_l c_l^2 \left[ \langle \tilde \varphi_{l} \varphi_{lr} \rangle_{v} - \langle \tilde \varphi_{l}\rangle_{v} \langle \varphi_{lr}\rangle_{v_r}  \right]
	\end{align}
\end{subequations}
Notice that $\Delta^{(0)}_{rs} = \frac{1}{K} \sum_l c_l^2 \, \left[ \langle \varphi_{rl} \varphi_{sl} \rangle_{v} - \langle \varphi_{rl}\rangle_{v} \langle \varphi_{sl}\rangle_{v} \right] = 0$ if $r\ne s$ since the Gaussian variable $v_{lr}$ is independent of $v_{ls}$.  
We then inserting~\eqref{eq::largeK_step1_reexp} back into~\eqref{eq::largeK_step1_first} and perform the integrals over $\hat h_r$ and $\hat v$; in general the integral is of the form 
\begin{equation}
	\begin{split}
		\label{eq::generic_identity}
		&\int \prod_r \frac{d h_r d \hat h_r}{2\pi} \frac{d 
			\eta d \hat \eta }{2\pi} e^{i \hat h_r (h_r - M_r) + i \hat \eta  (\eta  - N)  - \frac{1}{2} \sum_{rs} \Delta_{rs} \hat h_r \hat h_s  - \frac{\Xi}{2} \hat \eta ^2 - \hat \eta  \sum_r \Omega_r \hat h_r} f(\{h_r\}, \eta) \\
		&\simeq \int \prod_r \frac{d h_r d \hat h_r}{2\pi} D\eta  \, e^{i \sum_r \hat h_r (h_r - M_r - \frac{\eta }{\sqrt{\Xi}} \Omega_r) - \frac{1}{2} \sum_{rs} \left( \Delta_{rs} - \frac{\Omega_r \Omega_s}{\Xi} \right) \hat h_r \hat h_s} f(\{h_r\}, N + \sqrt{\Xi} \eta ) \\
		&= \int D\eta  \prod_r Dh_r \, f\left( \left\{ M_r +  \frac{\Omega_r}{\sqrt{\Xi}} \, \eta  + \sum_s \left(\sqrt{A}\right)_{rs} h_s \right\}, N + \sqrt{\Xi} \eta  \right)
	\end{split}
\end{equation}
with $A_{rs} \equiv \Delta_{rs} - \frac{\Omega_r \Omega_s}{\Xi}$. Specializing this identity to our case, i.e. using $f(\left\{h_r\right\}, \eta ) = e^{-\beta \sum_r \ell_r(h_r)} \, \widetilde{\ell}(\eta)$ we get
\begin{equation}
	\begin{split}
		E(\gamma) &= \int \prod_l Dx_{rl} \frac{\int \prod_r D\lambda_r D s \, \tilde{\ell}\left( N^{(0)} + \sqrt{\Xi^{(0)}} s \right) \, e^{-\beta \sum_r \ell\left( M_r^{(0)} + \frac{\Omega_r^{(0)}}{\sqrt{\Xi^{(0)}}}s + \sum_s \left[ \sqrt{\Lambda^{(0)}} \right]_{rs}  \lambda_s \right)} }{ \prod_r \int D \lambda \, e^{-\beta \ell_r\left( M_r^{(0)} + \sqrt{\Delta^{(0)}_r} \lambda \right) } }
	\end{split}
\end{equation}
with $\Lambda_{rs}^{(0)} \equiv \Delta^{(0)}_r \delta_{rs} - \frac{\Omega_r^{(0)} \Omega_s^{(0)}}{\Xi^{(0)}}$. Notice that we have done the same steps also in the denominator of the fraction.

Finally we apply the central limit again to simplify the integrals over $x_l$ variables; concerning all the variance terms, i.e. $\Delta_r^{(0)}$, $\Xi^{(0)}$ and $\Omega_r^{(0)}$ we can trivially compute their mean with respect to $x_r$, their variance being subleading in $K$. The only new terms come from the variance of the mean terms, i.e. the parameters $M_r^{(0)}$ and $N^{(0)}$. We get a term of the type
\begin{equation}
	\begin{split}
		&\int \prod_{rl} Dx_{rl} \int \prod_r \frac{d h_r d \hat h_r}{2\pi} \frac{d v d \hat v}{2\pi} \, e^{i h_r \hat h_r + i v \hat v - i \hat h_r M_r^{(0)} - i \hat v N^{(0)}} f(\{h_r\}, v) \\
		&\simeq \int \prod_r \frac{d h_r d \hat h_r}{2\pi} \frac{d v d \hat v}{2\pi} e^{i \hat h_r (h_r - M) + i \hat v (v - N)  - \frac{1}{2} \sum_{rs} \Psi_{rs} \hat h_r \hat h_s  - \frac{T}{2} \hat v^2 - \hat v \sum_r U_r \hat h_r} f(\{h_r\}, v) \\
		&= \int Dx \prod_r Dy_r \, f\left( \left\{ M +  \frac{U_r}{\sqrt{T}} \, x + \sum_s \sqrt{H}_{rs} y_s \right\}, N + \sqrt{T} x \right)
	\end{split}
\end{equation}
where we have used in the last step the general identity~\eqref{eq::generic_identity}. 
The final result is
\begin{equation}
	\begin{split}
		E(\gamma)
		&= \int \prod_l Dx_{rl} \frac{\int \prod_r D\lambda_r D s \, \tilde{\ell}\left( N^{(0)} + \sqrt{\Xi^{(0)}} s \right) \, e^{-\beta \sum_r \ell\left( M_r^{(0)} + \frac{\Omega_r^{(0)}}{\sqrt{\Xi^{(0)}}}s + \sum_s \left[ \sqrt{\Lambda^{(0)}} \right]_{rs}  \lambda_s \right)} }{ \prod_r \int D \lambda \, e^{-\beta \ell_r\left( M_r^{(0)} + \sqrt{\Delta^{(0)}_r} \lambda \right) } } \\
		&= \int Dx \prod_r Dy_r \frac{\int \prod_r D\lambda_r D s \, \tilde{\ell}\left( \sqrt{T} x + \sqrt{\Xi} s \right) \, e^{-\beta \sum_r \ell_r\left( \frac{U_r}{\sqrt{T}} x + \sum_s \sqrt{H}_{rs} y_s + \frac{\Omega_r}{\sqrt{\Xi}} s + \sum_s \left[ \sqrt{\Lambda} \right]_{rs}  \lambda_s \right)}}{ \prod_r \int D\lambda \, e^{-\beta \ell_r\left( \frac{U_r}{\sqrt{T}} x + \sum_s \sqrt{H}_{rs} y_s + \sqrt{\Delta_r} \lambda \right)}}
	\end{split}
	\label{eq::largeK_final}
\end{equation}
where we have defined the quantities
\begin{subequations}
	\begin{align}
		H_{rs} &\equiv \Psi_{rs} - \frac{U_r U_s}{T} \\
		\Lambda_{rs} &\equiv \Delta_{r} \delta_{rs} - \frac{\Omega_r \Omega_s}{\Xi}
	\end{align}
\end{subequations}	
and
\begin{subequations}
	\label{eq::effective_order_parameters_raw}
	\begin{align}
		\Psi_{rs} &\equiv \langle \langle \varphi_r \rangle_{v} \langle \varphi_s \rangle_{v} \rangle_x - \langle \varphi_r \rangle_{v, x} \langle \varphi_s \rangle_{v, x}\\
		\Delta_{r} &\equiv \langle \varphi^2 \rangle_{v, x} - \langle \langle \varphi_r \rangle_{v}^2 \rangle_x \\
		T &\equiv \langle \langle \tilde \varphi \rangle^2_{v} \rangle_x - \langle \tilde \varphi \rangle_{v, x}^2\\
		\Xi &\equiv \langle \tilde \varphi^2 \rangle_{v, x} - \langle \langle \tilde \varphi \rangle^2_{v} \rangle_x \\
		U_r &\equiv \langle \langle \varphi_r \rangle_v \langle \tilde \varphi \rangle_v \rangle_x - \langle \varphi \rangle_{v, x} \langle \tilde \varphi \rangle_{v, x}\\ 
		\Omega_r &\equiv \langle \langle \tilde \varphi \varphi_{r} \rangle_{v} \rangle_x - \langle \langle \tilde \varphi \rangle_{v} \langle \varphi_r \rangle_{v} \rangle_x 
	\end{align}
\end{subequations}
Note that in doing the last step in equation~\eqref{eq::largeK_final} we have used the fact that $M = N = 0$ since they are both proportional to $\sum_l c_l$. 
We show in appendix~\ref{sec::effective_order_parameters} that the quantities above can be all written in terms of the following function that depends on the activation function $\varphi$
\begin{equation}
	\Delta_Q(q) \equiv \int Dx \left[ \int Dy \, \varphi\left( \sqrt{q} x + \sqrt{Q-q} y \right) \right]^2 
\end{equation}
as
\begin{subequations}
	\label{eq::effective_order_parameters}
	\begin{align}
		\Psi_{rs} &= \Delta_Q(t_{rs}) - \Delta_Q(0)\\
		\Delta_{r} &= \Delta_Q(Q) - \Delta_Q(t_{rr}) \\
		T &=  \Delta_Q\left(  \frac{\gamma^2 q_1 + (1-\gamma)^2 q_2 + 2\gamma (1-\gamma) p}{c_\gamma^2} \right) - \Delta_Q(0)\\
		\Xi &= \Delta_Q(Q) -  \Delta_Q\left(  \frac{\gamma^2 q_1 + (1-\gamma)^2 q_2 + 2\gamma (1-\gamma) p}{c_\gamma^2} \right)\\
		U_1 &= \Delta_Q\left(\frac{\gamma q_1 + (1-\gamma) p}{c_\gamma}\right) - \Delta_Q(0)\\
		U_2 &= \Delta_Q\left(\frac{(1-\gamma) q_2 + \gamma p}{c_\gamma}\right) - \Delta_Q(0)\\
		\Omega_1 &= 
		\Delta_Q\left(\frac{\gamma Q + (1-\gamma) p}{c_\gamma}\right) - \Delta_Q\left(\frac{\gamma q_1 + (1-\gamma) p}{c_\gamma}\right) \\
		\Omega_2 &= 
		\Delta_Q\left(\frac{(1-\gamma)Q + \gamma p}{c_\gamma}\right) - \Delta_Q\left(\frac{(1-\gamma) q_2 + \gamma p}{c_\gamma}\right)
	\end{align}
\end{subequations}
We can simplify~\eqref{eq::largeK_final} by performing 2-dimensional rotations of the integration over the $\lambda$ variables
\begin{equation}
	\label{eq::final_result_theta}
	\begin{split}
		E(\gamma) 
		& =  \int Dx Dy_1 Dy_2 D s \,  \tilde{\ell}\left( \frac{U_1}{\sqrt{\Psi_{11}}} x + \frac{U_2 \Psi_{11} - U_1 \Psi_{12}}{\sqrt{\Psi_{11} \det \Psi}} y_1 + \sqrt{\frac{T \det H}{\det \Psi}} y_2 + \sqrt{\Xi} s \right) \\
		&\frac{ \int D\lambda_1 D\lambda_2 \, e^{-\beta \ell_1\left( \sqrt{\Psi_{11}} x + \frac{\Omega_1}{\sqrt{\Xi}} s + \sqrt{\Lambda_{11}} \lambda_1 \right)} e^{-\beta \ell_2\left( \frac{\Psi_{12}}{\sqrt{\Psi_{11}}} x + \sqrt{\frac{\det \Psi}{\Psi_{11}} } y_1 + \frac{\Omega_2}{\sqrt{\Xi}} s + \frac{\Lambda_{12}}{\sqrt{\Lambda_{11}}}\lambda_1 + \sqrt{\frac{\det \Lambda}{\Lambda_{11}}}\lambda_2 \right)}}{ \int D\lambda \, e^{-\beta \ell_1\left( \sqrt{\Psi_{11}} x + \sqrt{\Delta_1} \lambda \right)} \int D\lambda \, e^{-\beta \ell_2\left(  \frac{\Psi_{12}}{\sqrt{\Psi_{11}}} x + \sqrt{ \frac{\det \Psi}{\Psi_{11}} } y_1 +  \sqrt{\Delta_2} \lambda \right)}}
	\end{split}
\end{equation}
Notice how the variable $y_2$ appears only in $\tilde \ell$. Performing a rotation over $\lambda_1$ and $s$ 
\begin{equation}
	\begin{split}
		\label{eq::final_result_generic_theta}
		\epsilon_t(\gamma) 
		& =  \int Dx Dy_1 Dy_2 \int D\lambda_1 D s \,  \tilde{\ell}\left( \frac{U_1}{\sqrt{\Psi_{11}}} x + \frac{U_2 \Psi_{11} - U_1 \Psi_{12} }{\sqrt{\Psi_{11} \det\Psi }} y_1 + \sqrt{\frac{T \det H}{\det \Psi}} y_2 + \frac{\Omega_1}{\sqrt{\Delta_1}} \lambda_1 + \sqrt{\frac{\Lambda_{11} \Xi}{\Delta_1}}s \right) \\
		&\frac{ e^{-\beta \ell_1\left( \sqrt{\Psi_{11}} x + \sqrt{\Delta_1} \lambda_1 \right)} \int D\lambda_2 \, e^{-\beta \ell_2\left( \frac{\Psi_{12}}{\sqrt{\Psi_{11}}} x + \sqrt{ \frac{\det \Psi}{\Psi_{11}} } y_1 + \sqrt{\frac{\Delta_1}{\Xi \Lambda_{11}}} \Omega_2 s + \sqrt{\frac{\det \Lambda}{\Lambda_{11}}}\lambda_2 \right)}}{ \int D\lambda \, e^{-\beta \ell_1\left( \sqrt{\Psi_{11}} x + \sqrt{\Delta_1} \lambda \right)} \int D\lambda \, e^{-\beta \ell_2\left(  \frac{\Psi_{12}}{\sqrt{\Psi_{11}}} x + \sqrt{ \frac{\det \Psi}{\Psi_{11}} } y_1 +  \sqrt{\Delta_2} \lambda \right)}}
	\end{split}
\end{equation}
Notice how $\lambda_1$ disappears from the argument of the second loss function $\ell_2$. Finally~\eqref{eq::final_result_theta} can be further simplified by letting appear the same argument in $\ell_2$ in the numerator and denominator. This can be obtained by performing a rotation over the Gaussian variables $s$ and $\lambda_2$. After one can integrate explicitly over $y_2$ obtaining
\begin{equation}
	\label{eq::final_result_generic}
	\begin{split}
		E(\gamma) 
		& =  \int Dx Dy_1 \int D\lambda_1 D\lambda_2 \,  \frac{ e^{-\beta \ell_1\left( \sqrt{\Psi_{11}} x + \sqrt{\Delta_1} \lambda_1 \right)}  \, e^{-\beta \ell_2\left( \frac{\Psi_{12}}{\sqrt{\Psi_{11}}} x + \sqrt{ \frac{\det \Psi}{\Psi_{11}} } y_1 + \sqrt{\Delta_2} \lambda_2 \right)}}{ \int D\lambda \, e^{-\beta \ell_1\left( \sqrt{\Psi_{11}} x + \sqrt{\Delta_1} \lambda \right)} \int D\lambda \, e^{-\beta \ell_2\left(  \frac{\Psi_{12}}{\sqrt{\Psi_{11}}} x + \sqrt{ \frac{\det \Psi}{\Psi_{11}} } y_1 +  \sqrt{\Delta_2} \lambda \right)}} \\
		& \int Ds \, \tilde{\ell}\left( \frac{U_1}{\sqrt{\Psi_{11}}} x + \frac{U_2 \Psi_{11} - U_1 \Psi_{12}}{\sqrt{\Psi_{11} \det \Psi}} y_1 + \frac{\Omega_1}{\sqrt{\Delta_1}} \lambda_1 +  \frac{\Omega_2}{\sqrt{\Delta_2}} \lambda_2 + \sqrt{\Xi - \frac{\Omega_1^2}{\Delta_1} - \frac{\Omega_2^2}{\Delta_2} + \frac{T \det H}{\det \Psi}} s \right)
	\end{split}
\end{equation}

\subsection{Analytical predictions in some particular cases}
In the following we will specialize equation~\eqref{eq::final_result_generic} to several interesting cases depending on the Boltzmann distribution through which the endpoints $\gamma = 0, 1$ are sampled.

\subsubsection{Error counting loss with a margin}

\begin{figure}[t]
	\includegraphics[width=0.49\linewidth]{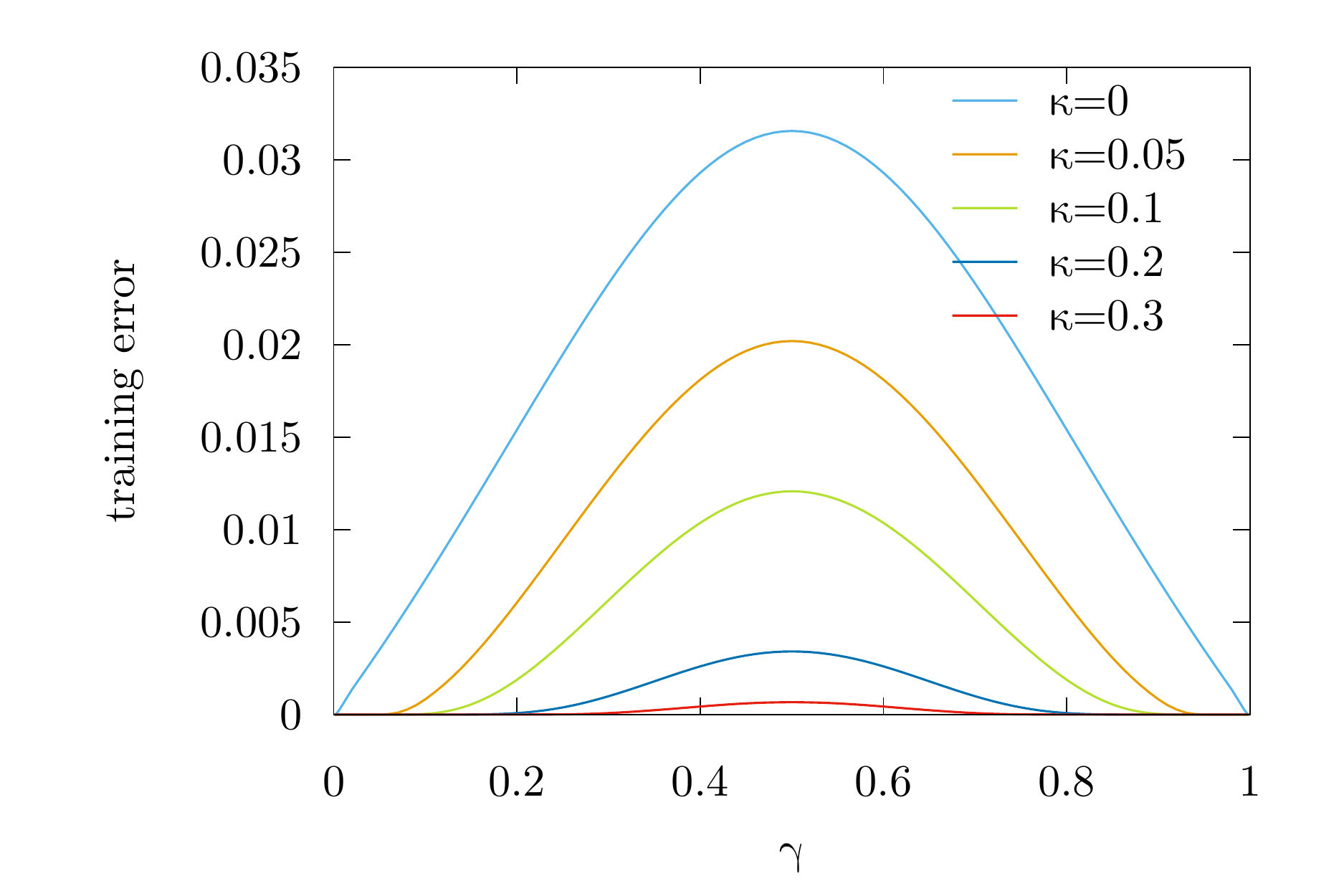}
	\includegraphics[width=0.49\linewidth]{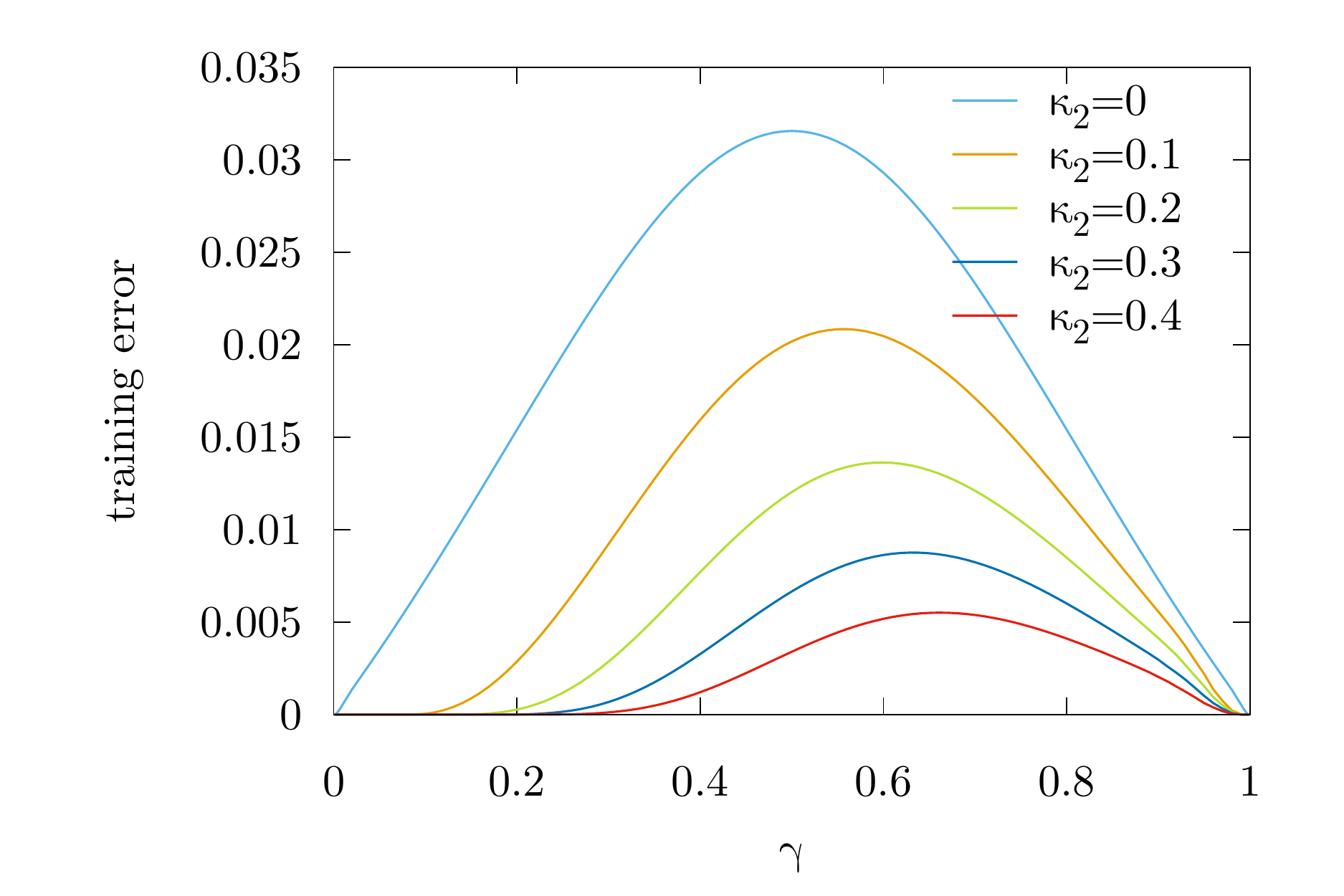}
	\caption{Training error on the geodesic paths among differently sampled solutions of the erf activation function with fixed squared norm $Q=1$. Left panel: Training error along the geodesic path connecting 2 solutions sampled from the loss function $\ell_r(x) \equiv \Theta(\kappa_r - x)$, $r = 1, 2$ with equal margin $\kappa_1 = \kappa_2 \equiv \kappa$ and $\alpha = 0.1$. The barrier between typical solutions i.e. $\kappa = 0$ is strictly non-vanishing; increasing the margin on the sampled solutions the barrier decreases and eventually vanishes for large enough $\kappa$. Right panel: Case where $\ell_r(x) \equiv \Theta(\kappa_r - x)$, $r = 1, 2$ and $\widetilde{\ell}(x) = \Theta(- x)$, with $\alpha = 0.1$. We here fixed the first endpoint (that is located at $\gamma = 1$) to have a margin $\kappa_1 = 0$. The second endpoint ($\gamma = 0$) has a margin $\kappa_2= 0, 0.1, 0.2, 0.3, 0.4$ (curves from top to bottom). As observed in the main text, increasing the robustness $\kappa_2$ the training error barrier on the geodesic monotonically decreases. For large enough $\kappa_2$ the solutions become geodesically connected. 
	}
	\label{fig::training_error_atypical_erf}
\end{figure}

Here we specialize~\eqref{eq::final_result_generic} to the case $\ell_1(x) = \Theta(-x + \kappa_1)$ and $\ell_2(x) = \Theta(-x + \kappa_2)$ where $\kappa_1\,, \kappa_2 \ge 0$ impose a certain degree of robustness $\boldsymbol{w}_1$ and $\boldsymbol{w}_2$. We further consider $\tilde \ell (x) = \Theta(-x)$ and we will focus on the $\beta\to \infty$ limit for simplicity. Starting from~\eqref{eq::final_result_theta} we have
\begin{equation}
	\begin{split}
		\epsilon_t(\gamma) 
		& =  \int Dx Dy_1 \, \\
		&\times \frac{ \int D s  H\left(\frac{ \frac{U_1}{\sqrt{\Psi_{11}}} x + \frac{U_2 \Psi_{11} - U_1 \Psi_{12}}{\sqrt{\Psi_{11} \det \Psi }} y_1 + \sqrt{\Xi} s }{\sqrt{\frac{T \det H}{\det \Psi}}}\right)  \int_{\frac{\kappa_1 - \sqrt{\Psi_{11}} x - \frac{\Omega_1}{\sqrt{\Xi}} s}{\sqrt{\Lambda_{11}}}}^{\infty} D\lambda_1 \, H\left(\frac{ \sqrt{\Lambda_{11}} \left(\kappa_2 - \frac{\Psi_{12}}{\sqrt{\Psi_{11}}} x - \sqrt{ \frac{\det \Psi}{\Psi_{11}}} y_1 - \frac{\Omega_2}{\sqrt{\Xi}} s \right) - \Lambda_{12}\lambda_1 }{\sqrt{\det \Lambda } }\right)    }{ H\left( \frac{\kappa_1 - \sqrt{\Psi_{11}}x }{\sqrt{\Delta_1}}\right) H\left( \frac{\kappa_2 - \frac{\Psi_{12}}{\sqrt{\Psi_{11}}} x - \sqrt{ \frac{\det \Psi}{\Psi_{11}} } y_1}{\sqrt{\Delta_2}}\right) }
	\end{split}
\end{equation}
This is the expression that we have used to produce the plots in Figure~\ref{fig::training_error_typical} of the main text. We show similar plots for the Erf activation function in Figure~\ref{fig::training_error_atypical_erf}. 
If the endpoints are sampled with the same margin $\kappa_1 = \kappa_2 = \kappa$ then as stated before $q_1 = q_2 = p$ and this implies the relation $\sqrt{\frac{T \det H}{\det \Psi}} = \sqrt{T - \frac{U^2}{\Psi}}$. In this case, the formula simplifies even further and reads
\begin{equation}
	\begin{split}
		\epsilon_t(\gamma) 
		&= \int Dx \frac{\int D s \, H\left( \frac{\frac{U}{\sqrt{\Psi}} x + \sqrt{\Xi} s}{\sqrt{T - \frac{U^2}{\Psi}}} \right) \, \int_{ \frac{\kappa -  \sqrt{\Psi} x - \frac{\Omega_1}{\sqrt{\Xi}} s }{\sqrt{\Delta - \frac{\Omega_1^2}{\Xi}}}}^\infty D\lambda_1    \, H \left( \frac{ \sqrt{\Delta - \frac{\Omega_1^2}{\Xi}} \left(\kappa - \sqrt{\Psi} x - \frac{\Omega_2}{\sqrt{\Xi}} s \right) + \frac{\Omega_1 \Omega_2}{\Xi} \lambda_1 }{\sqrt{\det \Lambda}} \right)}{H^2\left(\frac{ \kappa - \sqrt{\Psi} x}{\sqrt{\Delta}} \right)} \,.
	\end{split}
\end{equation}

\subsubsection{Equal general loss functions at finite temperature}

We will here suppose that $\ell_1 = \ell_2 = \tilde \ell \equiv \ell$ and $\beta < \infty$. In this case $q_1 = q_2 = p \equiv q$, so that $\Psi_{rs} = \Delta_Q(q) - \Delta_Q(0) \equiv \Psi$, $U_1 = U_2 \equiv U = \Delta_Q\left(\frac{q}{c_\gamma}\right) - \Delta_Q(0)$ and $\Delta_1 = \Delta_2 \equiv \Delta = \Delta_Q(Q) - \Delta_Q(q)$. In this case we obtain
\begin{equation}
	\label{eq::anal_result_main}
	\begin{split}
		E(\gamma) 
		& =  \int Dx  D\lambda_1 D\lambda_2 \,  \frac{ e^{-\beta \ell\left( \sqrt{\Psi} x + \sqrt{\Delta} \lambda_1 \right)}  \, e^{-\beta \ell\left( \sqrt{\Psi} x + \sqrt{\Delta} \lambda_2 \right)}}{ \left[\int D\lambda \, e^{-\beta \ell\left( \sqrt{\Psi} x + \sqrt{\Delta} \lambda \right)} \right]^2 } \\
		& \int Ds \, \ell\left( \frac{U}{\sqrt{\Psi}} x + \frac{\Omega_1}{\sqrt{\Delta}} \lambda_1 +  \frac{\Omega_2}{\sqrt{\Delta}} \lambda_2 + \sqrt{\Xi - \frac{\Omega_1^2}{\Delta} - \frac{\Omega_2^2}{\Delta} + T - \frac{U^2}{\Psi}} s \right)
	\end{split}
\end{equation}
The previous equation is the one that we have used for the theoretical curve presented in Figure~\ref{fig::comparison} of the main text. 
If one is interested in measuring the training error we have
\begin{equation}
	\begin{split}
		E(\gamma) 
		& =  \int Dx  D\lambda_1 D\lambda_2 \,  \frac{ e^{-\beta \ell\left( \sqrt{\Psi} x + \sqrt{\Delta} \lambda_1 \right)}  \, e^{-\beta \ell\left( \sqrt{\Psi} x + \sqrt{\Delta} \lambda_2 \right)}}{ \left[\int D\lambda \, e^{-\beta \ell\left( \sqrt{\Psi} x + \sqrt{\Delta} \lambda \right)}\right]^2 } \, H\left( \frac{\frac{U}{\sqrt{\Psi}} x + \frac{\Omega_1}{\sqrt{\Delta}} \lambda_1 +  \frac{\Omega_2}{\sqrt{\Delta}} \lambda_2}{\sqrt{\Xi - \frac{\Omega_1^2}{\Delta} - \frac{\Omega_2^2}{\Delta} + T - \frac{U^2}{\Psi}}} \right)
	\end{split}
\end{equation}
where $H(x) \equiv \frac{1}{2} \text{Erfc}\left( \frac{x}{\sqrt{2}} \right)$. In the previous formulas $T$ reduces to $T=\Delta_Q\left(  \frac{q}{c_\gamma^2} \right) - \Delta_Q(0)$.

\subsubsection{Generic loss -- error counting loss with a margin}\label{sec::xent_errorcounting}

\begin{figure}[t]
	\includegraphics[width=0.49\linewidth]{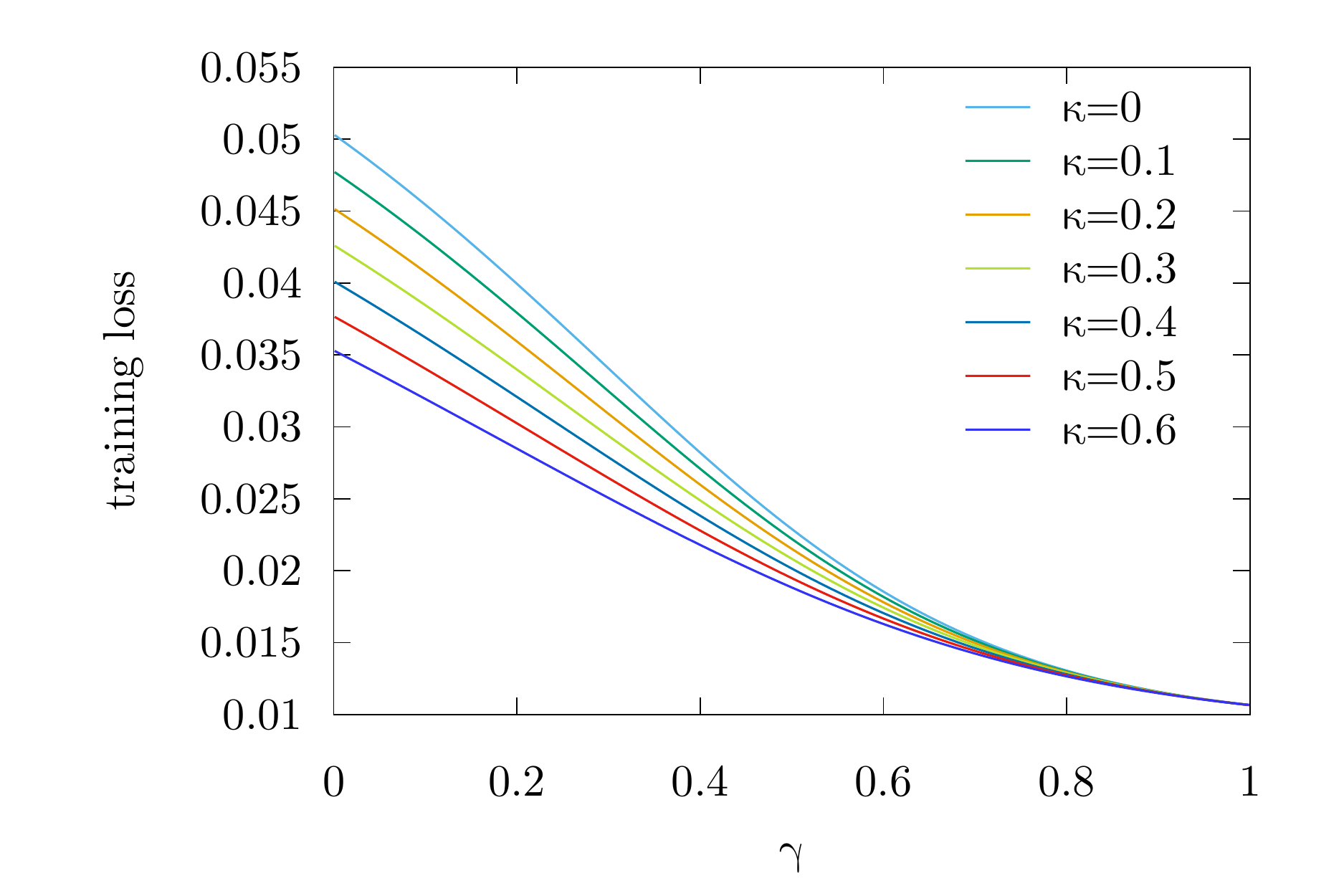}
	\includegraphics[width=0.49\linewidth]{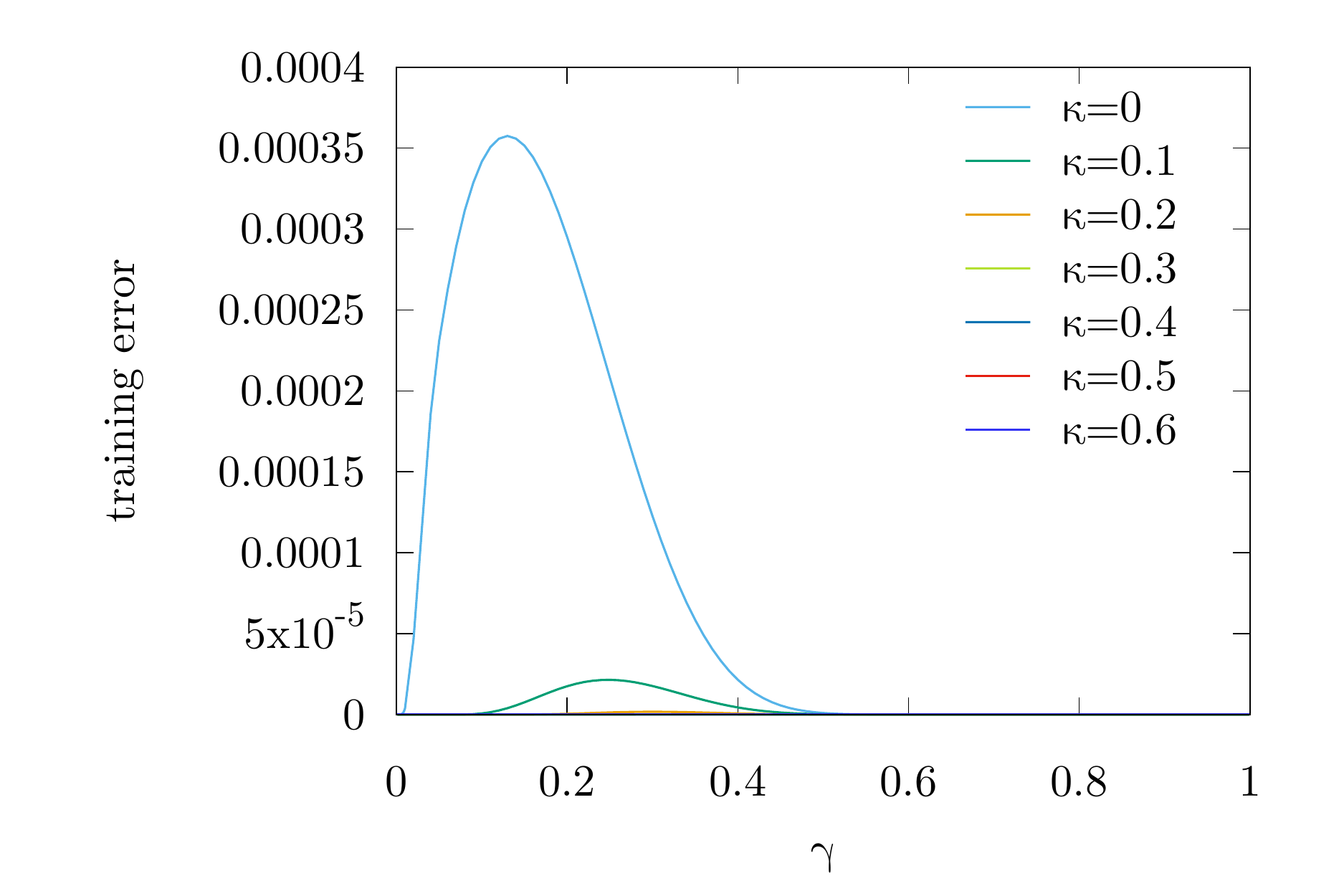}
	\caption{
		Training loss (left panel) and training error (right) on the geodesic path connecting a solution extracted from the loss function $\ell_2(x) = \Theta(-x + \kappa)$ (located at $\gamma=0$) and a solution of the cross entropy loss at $\beta \to \infty$ (located at $\gamma = 1$). Both endpoints and the configurations on the path are at a fixed squared norm $Q=1$. Both observables are plotted in the ReLU activation function case and for several values of the margin $\kappa$. Despite the training error presents a very small barrier near $\gamma=0$, which is appreciable for very low margin $\kappa$, the training loss is remarkably larger there. This suggests that the minimizers of the cross entropy are located deep in the bulk of the solution space manifold~\cite{baldassi2020shaping}.   
	}
	\label{fig::loss_theta_loss}
\end{figure}

The last case we consider is $\ell_2(x) = \Theta(-x + \kappa)$, but $\ell_1(x)$ is considered to be a generic convex loss function. We will again consider the infinite $\beta$ limit. This imposes a scaling on the overlap $q_1$ that reads
\begin{equation}
	q_1 = 1 - \frac{\delta q_1}{\beta}
\end{equation}
This induces a non-trivial scaling on some of the effective order parameters~\eqref{eq::effective_order_parameters}, in particular $\Delta_1$ and $\Omega_1$ will be vanishingly small
\begin{subequations}
	\begin{align}
		\Delta_1 &= \Delta_Q(Q) - \Delta_Q(q_1) \simeq \frac{\Delta_Q'(Q) \delta q_1}{\beta} \\
		\Omega_1 &= \Delta_Q\left(\frac{\gamma Q + (1-\gamma) p}{c_\gamma}\right) - \Delta_Q\left(\frac{\gamma q_1 + (1-\gamma) p}{c_\gamma}\right) \simeq  \Delta_Q'\left(\frac{\gamma Q + (1-\gamma) p}{c_\gamma}\right) \frac{\gamma \delta q_1}{c_\gamma \beta} \equiv \delta \Omega_1 \frac{\delta q_1}{\beta}
	\end{align}
\end{subequations} 
where we have introduced the quantity
\begin{equation}
	\Delta_Q'(q) \equiv  \frac{\partial \Delta_Q}{\partial q} = \int Dx \left[ \int Dy \, \varphi'\left(\sqrt{q} x + \sqrt{Q-q} y\right) \right]^2\,.
\end{equation}
and 
\begin{equation}
	\delta \Omega_1 \equiv  \Delta_Q'\left(\frac{\gamma Q + (1-\gamma) p}{c_\gamma}\right) \frac{\gamma}{c_\gamma}
\end{equation}
Furthermore, we have that $\Psi_{11} = \Delta_Q(Q)- \Delta_Q(0)$, $U_1 = \Delta_Q\left(\frac{\gamma Q + (1-\gamma) p}{c_\gamma}\right) - \Delta_Q(0)$,  $\frac{\Lambda_{11}}{\Delta_1} \to 1$ and $\Lambda_{12} \to 0$ so that $\frac{\det \Lambda}{\Lambda_{11}} \to \Lambda_{22}$. Using those relations inside~\eqref{eq::final_result_generic_theta}, rescaling $\lambda_1 \to \beta \lambda_1$ and using a saddle point over $\lambda_1$, one gets
\begin{equation}
	\begin{split}
		\epsilon_t(\gamma) 
		& =  \int Dx Dy_1 Dy_2 \int D s \,  \tilde{\ell}\left( \frac{U_1}{\sqrt{\Psi_{11}}} x + \frac{U_2 \Psi_{11} - U_1 \Psi_{12}}{\sqrt{\Psi_{11} \det \Psi}} y_1 + \sqrt{\frac{T \det H}{\det \Psi }} y_2 + \frac{\delta\Omega_1 \sqrt{\delta q_1}}{\sqrt{\Delta_Q'(Q)}} z_\star(x) + \sqrt{\Xi}s \right) \\
		&\times \frac{ H\left( \frac{\kappa - \frac{\Psi_{12}}{\sqrt{\Psi_{11}}} x - \sqrt{ \frac{\det \Psi}{\Psi_{11}} } y_1 - \frac{\Omega_2}{\sqrt{\Xi}} s}{\sqrt{\Lambda_{22}}} \right) }{ H\left( \frac{\kappa - \frac{\Psi_{12}}{\sqrt{\Psi_{11}}} x - \sqrt{  \frac{\det \Psi}{\Psi_{11}}} y_1}{\sqrt{\Delta_2}} \right)}
	\end{split}
\end{equation}
where $z_\star(x)$ is the function defined in~\eqref{eq::zstar} that also appears in the equilibrium computation~\cite{Baldassi2023Typical}. In the case we want to measure the training error, i.e. $\tilde \ell(x) = \Theta(-x)$ we have
\begin{equation}
	\begin{split}
		\epsilon_t(\gamma) 
		& =  \int Dx Dy_1 D s \,  H\left( \frac{\frac{U_1}{\sqrt{\Psi_{11}}} x + \frac{U_2 \Psi_{11} - U_1 \Psi_{12}}{\sqrt{\Psi_{11} \det \Psi}} y_1 + \frac{\delta\Omega_1 \sqrt{\delta q_1}}{\sqrt{\Delta_Q'(Q)}} z_\star(x) + \sqrt{\Xi}s}{\sqrt{\frac{T \det H}{\det \Psi}}} \right) \frac{ H\left( \frac{\kappa - \frac{\Psi_{12}}{\sqrt{\Psi_{11}}} x - \sqrt{ \frac{\det \Psi}{\Psi_{11}} } y_1 - \frac{\Omega_2}{\sqrt{\Xi}} s}{\sqrt{\Lambda_{22}}} \right) }{ H\left( \frac{\kappa - \frac{\Psi_{12}}{\sqrt{\Psi_{11}}} x - \sqrt{ \frac{\det \Psi}{\Psi_{11}} } y_1}{\sqrt{\Delta_2}} \right)}
	\end{split}
\end{equation}
We show in Figure~\ref{fig::loss_theta_loss} the training loss and error along the geodesics connecting solutions extracted from the error counting loss with a margin and the typical cross-entropy minimizer. Despite the training error is very small along the path, the loss is much larger in the neighborhood of the endpoint corresponding to the solution extracted from the error counting loss with a margin. As the margin is increased the training loss decreases.

\section{Effective order parameters} ~\label{sec::effective_order_parameters}

In this section we show that the effective order parameters defined in Eq.~\eqref{eq::effective_order_parameters_raw} reduce to the expressions given in~\eqref{eq::effective_order_parameters}. 

We remind the notation
\begin{subequations}
	\begin{align}
		\varphi_{r}(v_r, x) &\equiv \varphi \left(\sqrt{Q-q_r} v_{r} - \sum_{s} \mathcal{T}_{rs} x_{s} \right) \\
		\tilde{\varphi}(\gamma; v_1, v_2, x_1, x_2) &\equiv \varphi\left(\sum_r \gamma_r \left( \frac{\sqrt{Q-q_r}v_{r} - \sum_s \mathcal{T}_{rs} x_{s}}{c_{\gamma}} \right) \right)
	\end{align}
\end{subequations}
Remember also that $\mathcal{T}$ is the square root matrix of the matrix $t_{rs} = q_r \delta_{rs} + (1-\delta_{rs}) p$ and therefore we have the following identities $q_1 = \mathcal{T}_{11}^2 + \mathcal{T}_{12}^2$, $q_2 = \mathcal{T}_{22}^2 + \mathcal{T}_{12}^2$, $p = \mathcal{T}_{12} (\mathcal{T}_{11} + \mathcal{T}_{22})$ and $(\mathcal{T}_{11} \mathcal{T}_{22} - \mathcal{T}_{12}^2)^2 = q_1 q_2 - p^2$. 

Let's start by analyzing the terms $\Psi_{rs} \equiv \langle \langle \varphi_r \rangle_{v} \langle \varphi_s \rangle_{v} \rangle_x - \langle \varphi_r \rangle_{v, x} \langle \varphi_s \rangle_{v, x}$ and $\Delta_{r} \equiv \langle \varphi^2 \rangle_{v, x} - \langle \langle \varphi_r \rangle_{v}^2 \rangle_x$, which only involve $\varphi_r$. First notice that
\begin{subequations}
	\begin{align}
		\langle \varphi_r \rangle_{v, x} &= \int Dy \, \varphi(\sqrt{Q} y) = \sqrt{\Delta_Q(0)} \,,\qquad r = 1, 2 \\
		\langle \varphi_r^2 \rangle_{v, x} &= \int Dy \, \varphi^2(\sqrt{Q} y) = \Delta_Q(Q) \,,\qquad r = 1, 2
	\end{align}
\end{subequations}
and
\begin{equation}
	\begin{split}
		\langle \langle \varphi_1 \rangle_{v} \langle \varphi_2 \rangle_{v} \rangle_x &= \int D x_1 D x_2 Dv_1 D v_2 \, \varphi \left(\sqrt{Q-q_1} v_{1} + \sum_{s} \mathcal{T}_{1s} x_{s} \right) \, \varphi \left(\sqrt{Q-q_2} v_{2} + \sum_{s} \mathcal{T}_{2s} x_{s} \right) \\
		&= \int D x_1 D x_2 Dv_1 D v_2 \, \varphi \left(\sqrt{Q-q_1} v_{1} + \sqrt{q_1} x_1 \right) \, \varphi \left(\sqrt{Q-q_2} v_{2} + \frac{p}{\sqrt{q_1}} x_1 + \sqrt{ q_2 - \frac{p^2}{q_1} } x_2 \right)\\
		&=\int D x_1 D x_2 \, \varphi \left(\sqrt{Q} x_1 \right) \, \varphi \left( \frac{p}{\sqrt{Q}} \, x_1 + \sqrt{Q-\frac{p^2}{Q}} x_2 \right) = \Delta_Q(p) \\
	\end{split}
\end{equation}
Similarly, if $r = s$ we have
\begin{equation}
	\begin{split}
		\langle \langle \varphi_r \rangle_{v} \langle \varphi_r \rangle_{v} \rangle_x = \langle  \langle \varphi_r \rangle_{v}^2 \rangle_x = \Delta_Q(q_r)
	\end{split}
\end{equation}	
so that $\Psi_{rs} = \Delta_Q(t_{rs}) - \Delta_Q(0)$ and $\Delta_r = \Delta_Q(Q) - \Delta_Q(t_{rr})$.

Secondly let's analyze the terms which contain only $\tilde{\varphi}$, i.e. $T \equiv \langle \langle \tilde \varphi \rangle^2_{v} \rangle_x - \langle \tilde \varphi \rangle_{v, x}^2$ and
$\Xi \equiv \langle \tilde \varphi^2 \rangle_{v, x} - \langle \langle \tilde \varphi \rangle^2_{v} \rangle_x$. The terms $\langle \tilde{\varphi} \rangle_{v, x}$ and $	\langle \tilde{\varphi}^2 \rangle_{v, x}$ are easy to analyze since the integrand involves a sum of 4 uncorrelated Gaussian variables, which is Gaussian. We therefore get
\begin{subequations}
	\begin{align}
		\langle \tilde{\varphi} \rangle_{v, x} &= \int Dy \, \varphi(\sqrt{Q} y) = \sqrt{\Delta_Q(0)}\,, \\
		\langle \tilde{\varphi}^2 \rangle_{v, x} &= \int Dy \, \varphi^2(\sqrt{Q} y) = \Delta_Q(Q)\,.
	\end{align}
\end{subequations}	
Finally 
\begin{equation}
	\begin{split}
		\langle \langle \tilde \varphi \rangle^2_{v} \rangle_x &= \int Dx_1 Dx_2 \left[ \int Dv_1 Dv_2 \, \varphi\left(\sum_r \gamma_r \left( \frac{\sqrt{Q-q_r}v_{r} + \sum_s \mathcal{T}_{rs} x_{s}}{c_{\gamma}} \right) \right) \right]^2 \\
		&= \int Dx \left[ \int Dy \, \varphi\left( \frac{\sqrt{Q-2Q\gamma(1-\gamma) - \gamma^2 q_1 - (1-\gamma)^2q_2}}{c_\gamma} y + \frac{\sqrt{\gamma^2 q_1 + (1-\gamma)^2 q_2 + 2\gamma (1-\gamma) p}}{c_\gamma} \, x \right) \right]^2 \\
		&= \Delta_Q\left(  \frac{\gamma^2 q_1 + (1-\gamma)^2 q_2 + 2\gamma (1-\gamma) p}{c_\gamma^2} \right)
	\end{split}
\end{equation}
The last computation concerns the correlations between the function $\varphi_r$ and $\tilde{\varphi}$, which appear in the variables $U_r \equiv \langle \langle \varphi_r \rangle_v \langle \tilde \varphi \rangle_v \rangle_x - \langle \varphi \rangle_{v, x} \langle \tilde \varphi \rangle_{v, x}$, 
$\Omega_r \equiv \langle \langle \tilde \varphi \varphi_{r} \rangle_{v} \rangle_x - \langle \langle \tilde \varphi \rangle_{v} \langle \varphi_r \rangle_{v} \rangle_x$. We need therefore to evaluate the following two quantities $\langle \langle \tilde \varphi \varphi_{r} \rangle_{v} \rangle_x$ and $\langle \langle \tilde \varphi \rangle_{v} \langle \varphi_r \rangle_{v} \rangle_x$, for $r=1, 2$. We are going to analyze the case $r=1$, the other can be obtained by symmetry. We have
\begin{equation}
	\begin{split}
		\langle \langle \tilde \varphi \varphi_{1} \rangle_{v} \rangle_x &= \int Dx_1 Dx_2 D v_1 Dv_2 \, \varphi\left(\sum_r \gamma_r \left( \frac{\sqrt{Q-q_r}v_{r} + \sum_s \mathcal{T}_{rs} x_{s}}{c_{\gamma}} \right) \right) \varphi \left(\sqrt{Q-q_1} v_{1} + \sum_{s} \mathcal{T}_{1s} x_{s} \right) \\
		&= \int Dx_1 Dx_2 D v_1 Dv_2 \, \varphi(x_1) \\
		&\times \varphi\left( \frac{\gamma Q + (1-\gamma) p}{\sqrt{Q} c_\gamma} x_1 - \frac{(1-\gamma) p \sqrt{Q-q_1}}{\sqrt{q_1 Q} c_\gamma} v_1 + \frac{(1-\gamma) \sqrt{Q-q_2}}{c_\gamma} v_2 + \frac{1-\gamma}{c_\gamma} \sqrt{q_2 - \frac{p^2}{q_1}} x_2\right) \\
		&= \int Dx_1 Dx_2 \, \varphi(\sqrt{Q} x_1) \varphi\left( \frac{\gamma Q + (1-\gamma) p}{\sqrt{Q} c_\gamma} x_1 + \frac{1-\gamma}{c_\gamma} \sqrt{Q-\frac{p^2}{Q}} x_2\right) = \Delta_Q\left( \frac{\gamma Q + (1-\gamma)p}{c_\gamma} \right)
	\end{split}
\end{equation}
Notice that this does not depend on either $q_1$ nor $q_2$. The case $r=2$ can be  obtained by sending $\gamma \to 1-\gamma$.  
Let's analyze the term $\langle \langle \tilde \varphi \rangle_{v} \langle \varphi_r \rangle_{v} \rangle_x$ in the $r=1$ case
\begin{equation}
	\begin{split}
		\langle \langle \tilde \varphi \rangle_{v} \langle \varphi_1 \rangle_{v} \rangle_x &= \int Dx_1 D x_2 Dv_1 Dv_2 Dv_3 \, \varphi \left(\sqrt{Q-q_1} v_{3} + \sum_{s} \mathcal{T}_{1s} x_{s} \right) \, \varphi\left(\sum_r \gamma_r \left( \frac{\sqrt{Q-q_r}v_{r} + \sum_s \mathcal{T}_{rs} x_{s}}{c_{\gamma}} \right) \right) \\
		&=\int Dx_1  Dx_2 \, \varphi(\sqrt{Q} x_1) \varphi\left( \frac{\gamma q_1 + (1-\gamma) p}{\sqrt{Q} c_\gamma} x_1 + \sqrt{Q-\frac{(\gamma q_1 + (1-\gamma) p)^2}{Qc_\gamma^2}} x_2 \right) \\
		&= \Delta_Q\left(\frac{\gamma q_1 + (1-\gamma) p}{c_\gamma}\right)
	\end{split}
\end{equation}
The case $r=2$ can be obtained by sending $\gamma \to 1-\gamma$ and $q_1 \to q_2$, i.e.
\begin{equation}
	\begin{split}
		\langle \langle \tilde \varphi \rangle_{v} \langle \varphi_2 \rangle_{v} \rangle_x = \Delta_Q\left(\frac{(1-\gamma) q_2 + \gamma p}{c_\gamma}\right)
	\end{split}
\end{equation}

\section{Computing the overlap between differently sampled solutions} \label{sec::overlap_p}

The scope of this section is to find the typical overlap between two configurations $\boldsymbol{w}_1$ and $\boldsymbol{w}_2$ that are sampled from two (in principle different) distribution $p_1(\bullet \,; \mathcal{D})$ and $p_2(\bullet \,; \mathcal{D})$, see the definition in equation~\eqref{eq::p1_p2}. A way of computing this overlap has been sketched in~\cite{Annesi2023star}. Here we adopt a different approach, based on the Franz-Parisi entropy~\cite{franz1995recipes}. The Franz-Parisi entropy is defined as the average log of the number of configurations $\boldsymbol{w}_2 \sim p_2(\bullet; \mathcal{D})$ that are at a fixed overlap $p$ from the $\boldsymbol{w}_1 \sim p_1(\bullet; \mathcal{D})$. In formulas
\begin{subequations}
	\begin{align}
		\phi_{FP}(S) &\equiv \mathbb{E}_{\mathcal{D}} \int d \boldsymbol{w}_1 \, p_1(\boldsymbol{w}_1; \mathcal{D}) \ln \mathcal{N}_{\mathcal{D}}(\boldsymbol{w}_1; S) \\
		\mathcal{N}_{\mathcal{D}}(\boldsymbol{w}_1; S) &\equiv \int d \boldsymbol{w}_2 \, p_2(\boldsymbol{w}_2; \mathcal{D}) \, \delta\left( N S - \boldsymbol{w}_1 \cdot \boldsymbol{w}_2 \right)
	\end{align}
\end{subequations}
In the following we will call $\boldsymbol{w}_1$ the ``reference'' weight and $\boldsymbol{w}_2$ the ``slaved'' weight as it is constrained to stay at a distance given by the reference $\boldsymbol{w}_1$. In the following we will also suppose (as done in the main text), that the configuration $\boldsymbol{w}_2$ sampled from $p_2$, possesses the same squared norm $Q$ as the reference $\boldsymbol{w}_1$; this can be achieved by properly choosing the Lagrande multiplier $\lambda_2$.

The typical (i.e. the most probable) overlap is the one that maximizes the Franz-Parisi entropy
\begin{equation}
	p = \argmax_S \phi_{FP}(S)
\end{equation}

The Franz-Parisi can be computed with standard methods using a double replica trick. Here we refer to~\cite{Baldassi2023Typical,baldassi2021unveiling} for the derivation in the case of the perceptron. In the tree committee machine in the large width limit one gets
\begin{equation}
	\phi_{FP}(S) 
	= \text{extr}_{q_2, \, t} \left[\mathcal{G}_S(q_2, t, S) + \alpha \mathcal{G}_E(q_2, t, S) \right]
\end{equation}
where 
\begin{equation}
	\begin{split}
		\mathcal{G}_S 
		&= \frac{(Q^2 - S^2) (Q - 2 q_1 ) + Q q_1^2 - 2 q_1 S t + Q t^2}{2(Q-q_2)(Q-q_1)^2} +\frac{1}{2} \ln(2\pi) + \frac{1}{2}\ln\left(Q-q_2\right)
	\end{split}
\end{equation}
and 
\begin{subequations}
	\begin{align}
		\mathcal{G}_E 
		&= \int Dz_0 \, \frac{\int Dz_1 D z_2 \, e^{-\beta \ell_1\left(\sqrt{\Delta_Q(q_1) - \Delta_Q(0)} z_0 + \frac{\Delta_Q(S) - \Delta_Q(t)}{\sqrt{\Gamma}} z_1 + \sqrt{\eta} z_2 \right) } }{ \int Dz_1 \, e^{-\beta \ell_1\left(\sqrt{\Delta_Q(q_1) - \Delta_Q(0)} z_0 + \sqrt{\Delta_Q(Q) - \Delta_Q(q_1)} z_1 \right) } } \\
		&\times \ln \int D z_3 \, e^{-\beta \ell_2\left(\sqrt{\Delta_Q(Q) - \Delta_Q(q_2)} z_3 + \frac{\Delta_Q(t) - \Delta_Q(0)}{\sqrt{\Delta_Q(q_1) - \Delta_Q(0)}} z_0 + \sqrt{\Gamma} z_1\right)} \\
		\eta &\equiv \Delta_Q(Q) - \Delta_Q(q_1) - \frac{(\Delta_Q(S) - \Delta_Q(t))^2}{\Gamma} \\
		\Gamma &= \Delta_Q(q_2) - \Delta_Q(0) - \frac{(\Delta_Q(t) - \Delta_Q(0))^2}{\Delta_Q(q_1) - \Delta_Q(0)} 
	\end{align}
\end{subequations}
Notice that $q_1$ represents the typical overlap between reference configurations $\boldsymbol{w}_1 \sim p_1(\boldsymbol{w}_1; \mathcal{D})$.
Imposing that at the typical distance the Franz-Parisi presents a maximum we have 
\begin{equation}
	\frac{\partial \phi_{FP}}{\partial S} = 0 \implies
	\frac{Q S - 2 q_1S +q_1t}{(Q-q_2)(Q-q_1)^2} \simeq \frac{S}{(Q-q_1)(Q-q_2)} = \alpha \frac{\partial \mathcal{G}_E}{\partial S}  
\end{equation}
The first equality follows from the fact that at the maximum of the Franz-Parisi entropy one can verify that the saddle point equation impose $t = S$. One can then compute the right hand side explicitly, expanding the expression for $t \to S$. The typical overlap $p$ can be obtained finally by solving this implicit equation for $p$
	\begin{equation}
		\label{eq::typical_overlap_solutions}
		\begin{split}
			\frac{p}{\Delta_Q'(p) (Q-q_1)(Q-q_2)} &= \frac{\alpha}{\Delta_Q(p) - \Delta_Q(0)}\int Dz_0 \, \left[ \frac{\partial}{\partial z_0} \, \ln \int Dz_1 e^{-\beta \ell_1\left(\sqrt{\Delta_Q(q_1) - \Delta_Q(0)} z_0 + \sqrt{\Delta_Q(Q) - \Delta_Q(q_1)} z_1 \right)} \right] \\
			&\times \left[\int D z_1 \frac{\partial}{\partial z_0} \ln \int Dz_3 \, e^{-\beta \ell_2 \left(\sqrt{\Delta_Q(Q) - \Delta_Q(q_2)} z_3 + \frac{\Delta_Q(p) - \Delta_Q(0)}{\sqrt{\Delta_Q(q_1) - \Delta_Q(0)}} z_0 + \sqrt{\Gamma} z_1\right) } \right]
		\end{split}
	\end{equation}
	where we have introduced the quantity $\Delta_Q'(q)$ as in equation~\eqref{eq::effective_order_parameter_derivative} and we have redefined $\Gamma$ to be $\Gamma = \Delta_Q(q_2) - \Delta_Q(0) - \frac{(\Delta_Q(p) - \Delta_Q(0))^2}{\Delta_Q(q_1) - \Delta_Q(0)}$.
	Notice that equation~\eqref{eq::typical_overlap_solutions} depends non-trivially on $q_1$ and $q_2$ which represent respectively the typical overlap of configurations $\boldsymbol{w}_1$ and $\boldsymbol{w}_2$ that are extracted from $\boldsymbol{w}_1 \sim p_1(\bullet; \mathcal{D})$ and $\boldsymbol{w}_2 \sim p_2(\bullet; \mathcal{D})$ and that can be obtained by a standard equilibrium computation of the partition function in equation~\eqref{eq::partition_function}. In particular, when the solutions are sampled from the \emph{same distribution}, then $q_1 = q_2$, and one can verify that equation~\eqref{eq::typical_overlap_solutions} is trivially satisfied by $p=q_1=q_2$. 
	
	In the following subsection we specialize equation~\eqref{eq::typical_overlap_solutions} to several interesting sub-cases, in the large $\beta$ limit. 
	
	\subsection{The error counting loss with a margin}
	
	In the case one is interested in the theta loss, i.e. $\ell_1(x) = \Theta(\kappa_1 - x)$ and $\ell_2(x) = \Theta(\kappa_2 - x)$ the integrals in~\eqref{eq::typical_overlap_solutions} inside the logs can be performed and in the infinite $\beta$ limit one gets
	\begin{equation}
		\label{eq::typical_overlap_solutions_theta_theta}
		\begin{split}
			\frac{p}{\Delta_Q'(p)(Q-q_1)(Q-q_2)} 
			&= \alpha \int Dz_0 D z_1\, \frac{GH\left( \frac{\kappa_1 + \sqrt{\Delta_Q(q_1) - \Delta_Q(0)} z_0 }{\sqrt{\Delta_Q(Q) - \Delta_Q(q_1)}} \right)  GH\left( \frac{\kappa_2 + \frac{\Delta_Q(p) - \Delta_Q(0)}{\sqrt{\Delta_Q(q_1) - \Delta_Q(0)}} z_0 + \sqrt{\Gamma} z_1 }{\sqrt{\Delta_Q(Q) - \Delta_Q(q_2)}} \right)}{\sqrt{(\Delta_Q(Q) - \Delta_Q(q_1)) (\Delta_Q(Q) - \Delta_Q(q_2))}} \,.
		\end{split}
	\end{equation}
	This expression reduces to the one of the perceptron case computed in~\cite{Annesi2023star} if one specializes it to the identity activation function $\varphi(x) = x$ where $\Delta_Q(q) = q$.

	\subsection{Large $\beta$ limit: generic loss -- error counting loss with a margin}
	
	We consider here that the reference solution is sampled from a generic convex loss function, whereas $\ell_2(x) = \Theta(\kappa_2 - x)$. In the large $\beta$ limit we have that $q_1$ scales as $q_1 = Q - \frac{\delta q_1}{\beta}$ and correspondingly $\Delta_Q(Q) - \Delta_Q(q_1) \simeq \frac{\Delta_Q'(Q) \delta q_1}{\beta}$. Scaling $z_1 \to \beta z_1$ and using saddle point method, we therefore have
	\begin{equation}
		\begin{split}
			\frac{p}{\Delta_Q'(p) (Q-q_2)} 
			&= \frac{\alpha \sqrt{\delta q_1}}{\sqrt{(\Delta_Q(Q) - \Delta_Q(q_2)) \Delta'_Q(Q) }}\int Dz_0 \, z_\star(z_0) \int D z_1 \, GH\left( \frac{\kappa_2 - \frac{\Delta_Q(p) - \Delta_Q(0)}{\sqrt{\Delta_Q(Q) - \Delta_Q(0)}} z_0 - \sqrt{\Gamma} z_1 }{\sqrt{\Delta_Q(Q) - \Delta_Q(q_2)}} \right)
		\end{split}
	\end{equation}
	where $z_\star(z_0)$ is the same function defined in~\eqref{eq::zstar}.

\end{document}